\UseRawInputEncoding
\documentclass[fleqn,usenatbib]
{mnras}
\usepackage{scalerel}
\usepackage{tikz}
\usetikzlibrary{svg.path}

\definecolor{orcidlogocol}{HTML}{A6CE39}
\tikzset{
  orcidlogo/.pic={
    \fill[orcidlogocol] svg{M256,128c0,70.7-57.3,128-128,128C57.3,256,0,198.7,0,128C0,57.3,57.3,0,128,0C198.7,0,256,57.3,256,128z};
    \fill[white] svg{M86.3,186.2H70.9V79.1h15.4v48.4V186.2z}
                 svg{M108.9,79.1h41.6c39.6,0,57,28.3,57,53.6c0,27.5-21.5,53.6-56.8,53.6h-41.8V79.1z M124.3,172.4h24.5c34.9,0,42.9-26.5,42.9-39.7c0-21.5-13.7-39.7-43.7-39.7h-23.7V172.4z}
                 svg{M88.7,56.8c0,5.5-4.5,10.1-10.1,10.1c-5.6,0-10.1-4.6-10.1-10.1c0-5.6,4.5-10.1,10.1-10.1C84.2,46.7,88.7,51.3,88.7,56.8z};
  }
}

\newcommand\orcidicon[1]{\href{https://orcid.org/#1}{\mbox{\scalerel*{
\begin{tikzpicture}[yscale=-1,transform shape]
\pic{orcidlogo};
\end{tikzpicture}
}{|}}}}

\usepackage{hyperref}

\usepackage{newtxtext,newtxmath}

\usepackage[T1]{fontenc}
\usepackage[title]{appendix}

\DeclareRobustCommand{\VAN}[3]{#2}
\let\VANthebibliography\thebibliography
\def\thebibliography{\DeclareRobustCommand{\VAN}[3]{##3}\VANthebibliography}

\usepackage{graphicx}	
\usepackage{amsmath}	

\newcommand{\be}{\begin{equation}}
\newcommand{\ee}{\end{equation}}

\newcommand{\msun}{M$_\odot$}

\newcommand{\mearth}{M$_\oplus$}

\title{\textcolor{black}{Sub-m s$^{-1}$ upper limits from} a deep HARPS-N radial-velocity search for planets orbiting HD\,166620 and HD\,144579\thanks{Results presented in this paper have been obtained within the HARPS-N Collaboration.}}

\author[Anna John et al.]{
A. Anna John \orcidicon{0000-0002-1715-6939},$^{1,2}$\thanks{E-mail: aaj1@st-andrews.ac.uk}
A. Collier Cameron \orcidicon{0000-0002-8863-7828},$^{1,2}$
J. P. Faria,$^{3}$
A. Mortier \orcidicon{0000-0001-7254-4363},$^{4}$
T. G. Wilson \orcidicon{0000-0001-8749-1962},$^{1,2}$
\newauthor
L. Malavolta,$^{11,12}$
L.A. Buchhave,$^{5}$
X. Dumusque,$^{8}$
M. L\'opez-Morales \orcidicon{0000-0003-3204-8183},$^{9}$
R.D. Haywood \orcidicon{0000-0001-9140-3574},$^{7,13}$
\newauthor
K. Rice \orcidicon{0000-0002-6379-9185},$^{6,14}$
A. Sozzetti \orcidicon{0000-0002-7504-365X},$^{10}$
D. W. Latham, $^{9}$
S. Udry,$^{15}$
F. Pepe, $^{20}$
M.\,Pinamonti, $^{10}$
A.\,Vanderburg, $^{21}$
\newauthor
A. Ghedina, $^{18}$
R.Cosentino,$^{18}$
M. Stalport, $^{19,20}$
B. A. Nicholson, $^{16,17}$
A. Fiorenzano \orcidicon{0000-0002-4272-4272},$^{18}$
E.\,Poretti $^{18}$
\\
$^{1}$SUPA, School of Physics \& Astronomy, University of St Andrews, North Haugh, St Andrews, KY169SS, UK\\
$^{2}$Centre for Exoplanet Science, University of St Andrews, North Haugh, St Andrews, KY169SS, UK\\
$^{3}$Instituto de Astrof\'isica e Ci\^encias do Espa\c{c}o, Universidade do Porto, CAUP, Rua das Estrelas, 4150-762 Porto, Portugal\\
$^{4}$School of Physics \& Astronomy, University of Birmingham, Edgbaston, Birmingham B15 2TT, UK\\
$^{5}$DTU Space, National Space Institute, Technical University of Denmark, Elektrovej 328, DK-2800 Kgs. Lyngby, Denmark\\
$^{6}$SUPA, Institute for Astronomy, University of Edinburgh, Blackford Hill, Edinburgh, EH9 3HJ, UK\\
$^{7}$Astrophysics Group, University of Exeter, Exeter EX4 2QL, UK\\
$^{8}$Observatoire Astronomique de l'Universit\'e de Gen\'eve, Chemin Pegasi 51, 1290 Versoix, Switzerland\\
$^{9}$Center for Astrophysics \textbar{} Harvard \& Smithsonian, 60 Garden Street, Cambridge, MA 02138, USA\\
$^{10}$INAF - Osservatorio Astrofisico di Torino,Via Osservatorio 20, I-10025 Pino Torinese, Italy\\
$^{11}$Dipartimento di Fisica e Astronomia "Galileo Galilei" - Università di Padova, Vicolo dell'Osservatorio 3, 35122, Padova, Italy; \\
$^{12}$INAF - Osservatorio Astronomico di Padova, Vicolo dell'Osservatorio 5, 35122, Padova, Italy\\
$^{13}$STFC Ernest Rutherford Fellow\\
$^{14}$Centre for Exoplanet Science, University of Edinburgh, Edinburgh, EH9 3FD, UK\\
$^{15}$Observatoire de Gen\'eve, Universit\'e de Gen\'eve, 51 chemin des Maillettes, 1290 Sauverny, Switzerland\\
$^{16}$University of Southern Queensland, Centre for Astrophysics, Toowoomba, Australia, 4350\\
$^{17}$Sub-department of Astrophysics, University of Oxford, Keble Rd, Oxford, United Kingdom, OX13RH\\
$^{18}$Fundaci\'{o}n Galileo Galilei -- INAF, Rambla Jos\'{e} Ana Fernandez P\'{e}rez 7, 38712 -- Bre\~{n}a Baja, Spain\\
$^{19}$Space sciences, Technologies and Astrophysics Research (STAR) Institute, Universit\'e de Li\`ege, All\`ee du 6 Ao\^ut 19C, 4000 Li\`ege, Belgium 
\\
$^{20}$D\'epartement d'Astronomie, Universit\'e de Gen\`eve, Chemin Pegasi 51b, CH-1290 Versoix, Suisse\\
$^{21}$Department of Physics and Kavli Institute for Astrophysics and Space Research, Massachusetts Institute of Technology, 77
Massachusetts Avenue, Cambridge, MA 02139, USA\\}
\date{Accepted XXX. Received YYY; in original form ZZZ}
\pubyear{2023}

\usepackage{fancyhdr}

\begin{document}

\label{firstpage}
\pagerange{\pageref{firstpage}--\pageref{lastpage}}

\maketitle

 \begin{abstract}
Minimising the impact of stellar variability in Radial Velocity (RV) measurements is a critical challenge in achieving the 10 cm s$^{-1}$ precision needed to hunt for Earth twins. Since 2012, a dedicated programme has been underway with HARPS-N, to conduct a blind RV Rocky Planets Search (RPS) around bright stars in the Northern Hemisphere. Here we describe the results of a comprehensive search for planetary systems in two RPS targets, HD\,166620 and HD\,144579.
Using wavelength-domain line-profile decorrelation vectors to mitigate the stellar activity and performing a deep search for planetary reflex motions using a trans-dimensional nested sampler, 
we found no significant planetary signals in the data sets of either of the stars.
We validated the results via data-splitting and injection recovery tests. Additionally, we obtained the 95$^{\rm {th}}$ percentile detection limits on the HARPS-N RVs. We found that the likelihood of finding a low-mass planet increases noticeably across a wide period range when the inherent stellar variability is corrected for using {\sc scalpels} \textbf{U}-vectors. We are able to detect planet signals with $M\sin i \leq 1$~\mearth for orbital periods shorter than 10 days. 
We demonstrate that with our decorrelation technique, we are able to detect signals as low as 54 cm s$^{-1}$, which brings us closer to the calibration limit of 50 cm s$^{-1}$ demonstrated by HARPS-N.
Therefore, we show that we can push down towards the RV precision required to find Earth analogues using high-precision radial velocity data with novel data-analysis techniques.

\end{abstract}




\section{Introduction}

\begin{figure}
    \centering
	\includegraphics[width=\columnwidth]{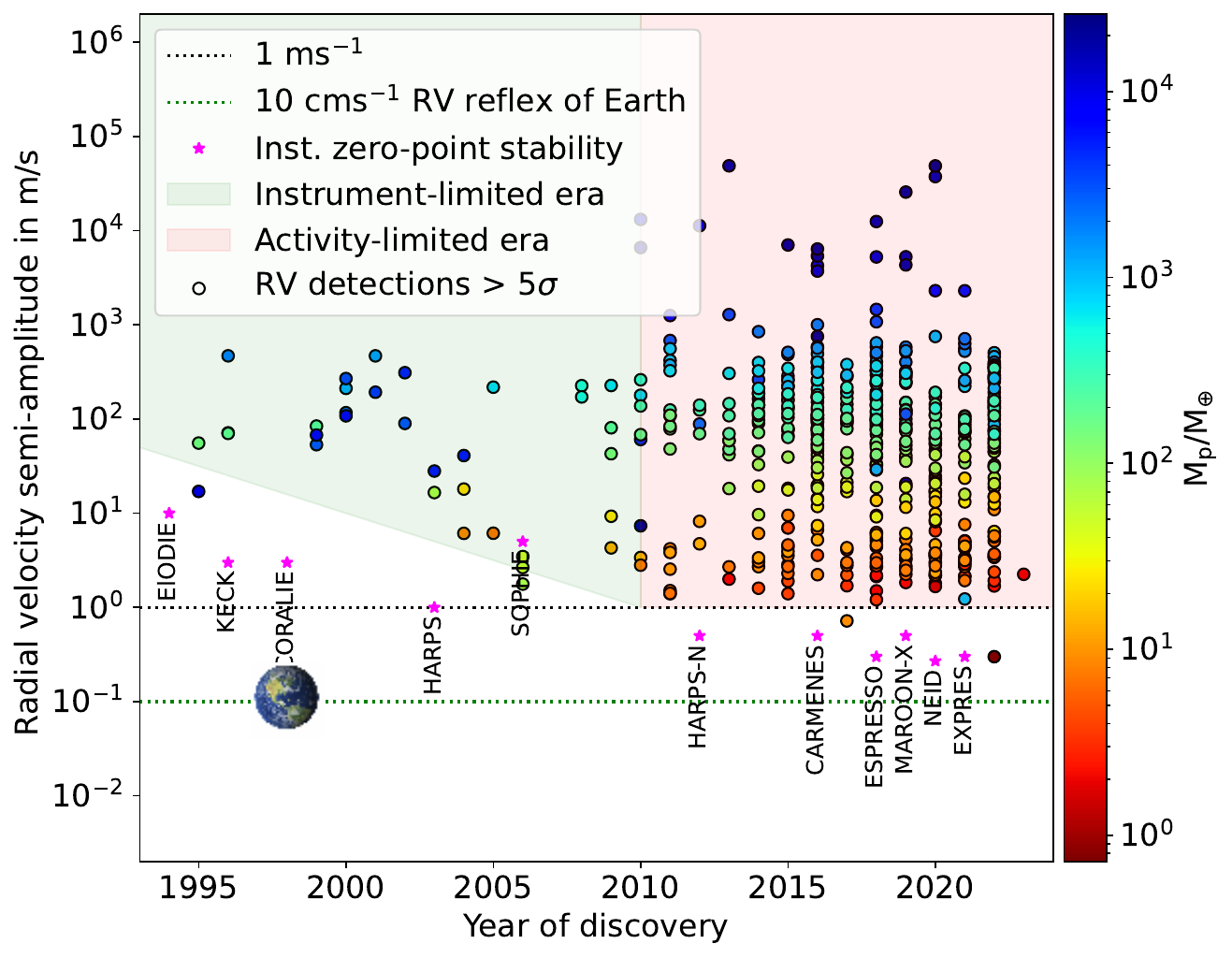}
    \caption{ The radial velocity semi-amplitudes of planets obtained from the NASA Exoplanet Archive are shown as a function of the discovery year. The green-shaded region shows an era in which the detection threshold improved almost exponentially in response to the improvement in wavelength calibration. However, this downward trend stops and flattens out in the early 2000s, despite the significant improvement in the instrument precision achieved by the new generation spectrographs. }
    \label{fig:submsbarrier}
\end{figure}

 A Radial Velocity (RV) precision of 0.1 m s$^{-1}$ is essential for the detection of Earth-sized low-mass exoplanets \textcolor{black}{orbiting in the habitable zones of Sun-like stars}.  Each subsequent advancement in instrumental stability since the discovery of the first exoplanet \textcolor{black}{around a Sun-like main sequence star} in 1995, has been reflected in a reduction in the detection limit of exoplanets down to 1 m s$^{-1}$ (green shaded region in Figure\,\ref{fig:submsbarrier}). Thanks to improvements in precision, it is now possible to use the new generation of spectrometers to observe the reflex motions of stars hosting such low-mass planets. However, the red-shaded region in Figure\,\ref{fig:submsbarrier} shows that despite these further instrumental advancements, the declining trend in the detectable RV semi-amplitude has hit a "1 m s$^{-1}$ saturation limit", at an amplitude comparable to RV variations produced by the intrinsic variability of the host star. 

Stellar variability is an umbrella term for a variety of astrophysical processes that have an impact on the RV of stars on timescales ranging from minutes to days. While stellar $p$-modes impose RV variability of up to 1 m s$^{-1}$ on a minutes scale in solar-type stars \citep[e.g.,][]{2019AJ....157..163C,2001A&A...374..733B} ), granulation introduces variability on timescales of hours to days \citep[e.g.,][]{2009ApJ...697.1032G,2015A&A...583A.118M}. Astronomers combat these parasitic signals by using longer exposure times and averaging out (for $p$-modes), and observing the stellar target multiple times throughout the night at intervals of several hours (for granulation) \citep{2011A&A...525A.140D}. On magnetically active stars, starspots and faculae/plage can cause RV shifts of the order of 1--100 m s$^{-1}$, while the suppression of convection causes shifts of about 10--20 m s$^{-1}$ \citep{2010A&A...512A..39M}.
Magnetic activity poses the biggest obstacle to the identification and characterization of small exoplanets, 
as it can occur on similar timescales and can completely obscure or mimic the RV signals of real exoplanets.

Recent studies have had some success modelling stellar variability by employing photometry to independently estimate the impact of stellar activity on the RVs \citep{2012MNRAS.419.3147A}. Other approaches include using time domain techniques like Gaussian Process (GP) regression to model the correlated noise induced by stellar activity \citep[e.g.,][]{2014MNRAS.443.2517H,2015MNRAS.452.2269R,2020A&A...638A..95D} and using multidimensional GP to use spectroscopic activity-indicators in conjunction with RVs to constrain the activity-induced signal in the RV time-series  \citep[e.g.,][]{2015MNRAS.452.2269R,2022MNRAS.509..866B}. 
Using neural networks to distinguish activity signals from true center-of-mass RV shifts of dynamical origin \citep{2021AAS...23733204D} and de-trending the RVs for line shape variations using the {\sc scalpels} basis vectors \citep{2021MNRAS.505.1699C} or spectral {\sc shell} vectors \citep{2022A&A...659A..68C} are a few wavelength domain techniques. Doppler imaging is also used to model both activity-induced variations and planet-induced shifts simultaneously in the spectral line profiles \citep{2022MNRAS.512.5067K}. 
A self-consistent comparison of these different methods (and more) is presented in \citet{2022AJ....163..171Z} .


In this paper, we present a deep search for planetary reflex-motion in the HARPS-N data of bright K \& G dwarfs HD\,166620 and HD\,144579, obtained as part of the Rocky Planet Search programme. We use {\sc tweaks} (\textbf{T}ime and \textbf{W}avelength-domain st\textbf{E}llar \textbf{A}ctivity mitigation using {\sc \textbf{K}ima} and {\sc \textbf{S}calpels}): an RV analysis  pipeline designed for attaining sub-m s$^{-1}$ detection thresholds at long orbital periods, by combining wavelength-domain and time-domain stellar activity mitigation \citep{2022MNRAS.515.3975A}.
This approach takes the posterior probability distributions for the orbital parameters of a system determined along with a complete Keplerian solver and a Gaussian Process (on/off) using the nested sampling algorithm, {\sc kima} \citep{2018JOSS....3..487F}. We then use {\sc scalpels} basis vectors for spectral line-shape decorrelation \citep{2021MNRAS.505.1699C}, following the result from \citet{2022MNRAS.515.3975A} which shows that de-trending the RVs for line shape variations using the {\sc scalpels} basis vectors yields a significantly better RV model over a model that does not account for these stellar activity signatures. There are also other studies that use {\sc scalpels} and {\sc tweaks} for improved detection of exoplanets (\citet{2022MNRAS.511.1043W}, Stalport 2023 submitted, Palethorpe 2023 in prep., Dalal 2023 in prep.).

\section{Targets \& observations}
\label{sec:Maunder minimum}
We applied {\sc tweaks} to two targets: HD\,166620 and HD\,144579. 
HD\,166620 and HD\,144579 have been observed for more than a decade with HARPS-N \citep{2012SPIE.8446E..1VC}, as part of the HARPS-N collaboration Rocky Planet Search (RPS) initiative.
The RPS programme was initiated in 2012 to search for small planets orbiting bright, nearby stars (mainly K and early M dwarfs) when the Kepler field was unobservable. The major objective of the RPS program is to conduct a systematic search for low-mass planets around nearby quiet stars that are visible from the northern hemisphere via extensive monitoring of the RV, with extremely high precision \citep{2015A&A...584A..72M}. Targets for this programme are monitored 2-3 times per night (to mitigate granulation), with exposure times of 15 minutes, \textcolor{black}{which is the cumulative open shutter time during multiple sub-exposures taken to prevent saturation of the brightest spectral orders}. This helps to average out the p-mode oscillations \citep{2019AJ....157..163C}, and thereby reduce the impact of stellar variability with brief typical timescales. 

\begin{figure}
    \centering
	\includegraphics[width=\columnwidth]{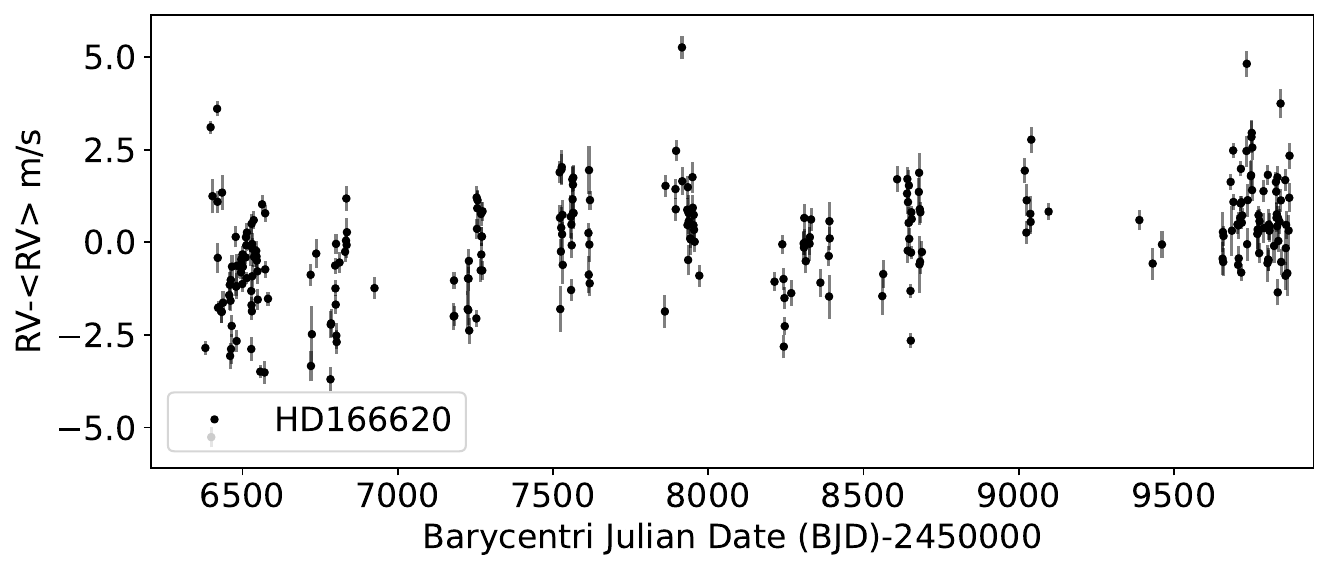}
	\includegraphics[width=\columnwidth]{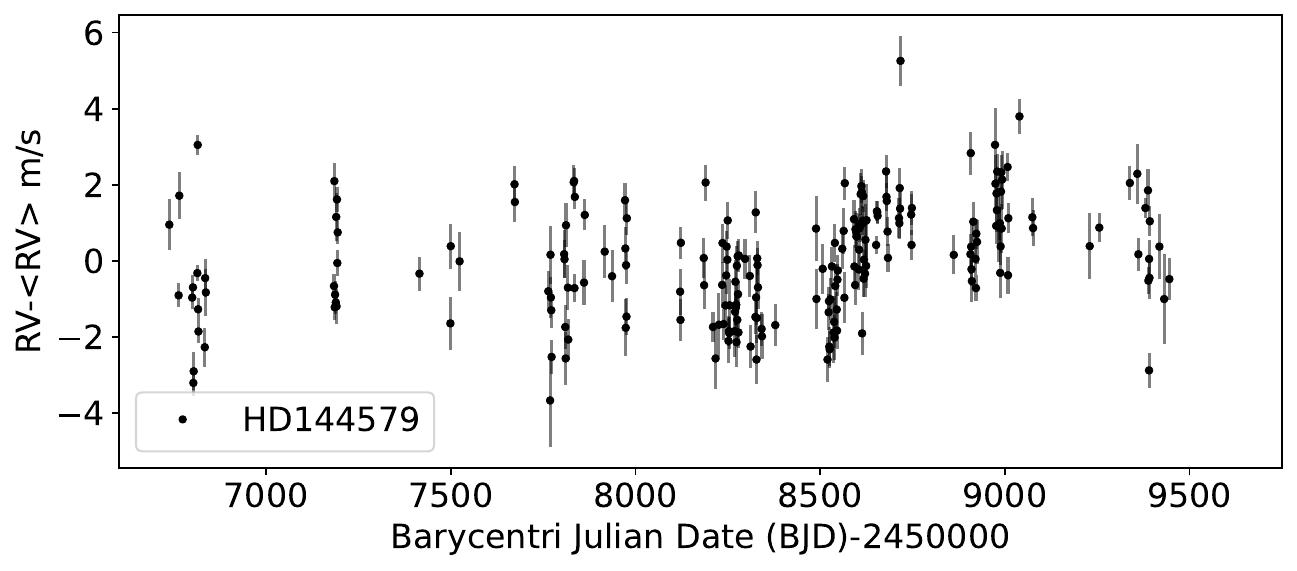}
    \caption{\textcolor{black}{Nightly binned} radial velocity observations (median-subtracted) for HD\,166620 and HD\,144579 measured from HARPS-N spectra are shown in the top and bottom panels, respectively.}
    \label{fig: Observations}
\end{figure}


A total of 1025 observations of HD\,166620 were obtained between 2013 March and 2023 March, and 888 observations of HD\,144579 were taken between 2013 February and 2022 March. 
For the mitigation of p-mode oscillations and granulation, we use the daily binned observations in this study, which reduces the total number of RV observations for HD\,166620 and HD\,144579 to 293 and 256, respectively (Figure\,\ref{fig: Observations}). After binning, the radial velocity RMS for HD\,166620 reduced from 1.8 to 1.45 m s$^{-1}$ and a reduction from 3.5 to 1.29 m s$^{-1}$ was observed for HD\,144579.

The HARPS-N Data Reduction Software (DRS), version 2.3.5 \citep{2021A&A...648A.103D}, was used to reduce the spectra used in this research. The DRS uses a mask (K2 \& G9 spectral types were used for HD\,166620 and HD\,144579,  respectively) composed of a multitude of lines that span the entire HARPS-N spectral range to calculate the Cross-Correlation Function (CCF), which is a representation of the average shape of the absorption profiles across the whole spectrum. Fitting a Gaussian to the CCFs then determines the stellar radial velocity, Full-width at Half Maximum (FWHM) and  CCF Area (product of FWHM and the central line depth). In this analysis, we use the CCFs as they are obtained from the DRS. The DRS pipeline also provides supplementary information such as the spectral bisector span (BIS), H$\alpha$, Na- Index, and Mount Wilson S-index for each observation. These additional data products are widely used as stellar activity indicators.

HD\,166620 is a K2V type star with ${\rm V_{mag} = 6.38}$ \citep{2015A&A...584A..72M}. 
The Mount Wilson program (1966--2001) \citep{1991ApJS...76..383D} studied the stellar chromospheric activity of HD\,166620, leaving us with an extensive record of the stellar chromospheric activity \citep{1995ApJ...438..269B}. \citet{2022AJ....163..183B} updated the stellar parameters for HD\,166620 and identified this star as a possible Maunder minimum candidate based on the five decades of Ca{\rm II} H\&K measurements from the Mount Wilson Survey , and continued observations at Keck as part of the California Planet Search programme \citep[][]{2004AJ....128.1273W, 2010ApJ...725..875I}. Recently, \citet{2022arXiv220700612L} presented additional Mount Wilson data that definitively traced the transition from cyclic activity to a prolonged phase of flat activity. 

\begin{figure}
    \centering
	
    \includegraphics[width=\columnwidth]{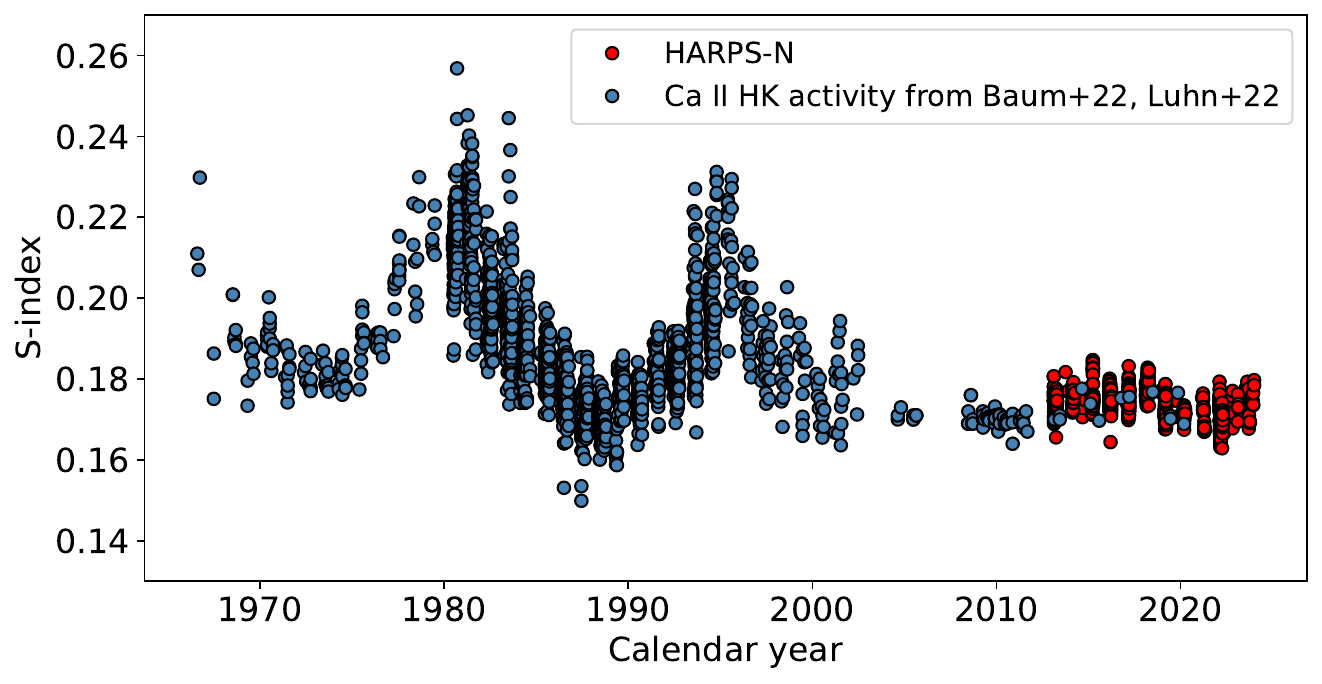}

    \caption{ The Ca II H\&K S-index data of HD\,166620 from Mount Wilson and Keck-Hires, displaying the transition into the Maunder Minimum  \citep[e.g.,][]{2022AJ....163..183B,2022arXiv220700612L}. The 10 years of S-index data obtained from HARPS-N, shown by the overlapping red points reinforces the star's present Maunder minimum phase. We adjusted for the offset in the HARPS-N data with a constant shift to have a consistent median S-index value with the Mount Wilson data set. Altogether we present 57 years of stellar chromospheric activity record for HD\,166620 here.} 
    \label{fig:Maunderminimum}
\end{figure}

\citet{2022AJ....163..183B} and \citet{2022arXiv220700612L} concluded HD\,166620 to be the first unambiguous Maunder-Minimum analogue, as shown by its S-index activity time series, as it changed from a cycling to a flat activity state.
We reproduce \citet{2022arXiv220700612L}'s data along with the 10 years of S-index data from HARPS-N here in Figure\,\ref{fig:Maunderminimum}.
This `flattened' state is explained by \citet{2022arXiv220700612L} as a less active phase of the star with fewer residual active areas to account for the even lower variability. This span of low activity includes the period of time from 2012 to the present, during which the radial velocity of HD\,166620 has been monitored intensively for planetary reflex-motion as  part of the HARPS-N Rocky Planet Search (RPS) programme.

The second target of this study is HD\,144579. In contrast to HD\,166620, no in-depth literature is presently available about this extremely bright G8V star with ${\rm V_{mag} = 6.65}$. The stellar parameters obtained as part of this study together with the Gaia Early Data Release \citep{2021A&A...650C...3G} for both the stars are given in Section \ref{subsec:stellarcharacteristics} and Table \ref{tab:HD166620parameters}.
\begin{table}
    
\caption{Stellar parameters obtained for HD\,166620 and HD\,144579 from three independent methods. Our final adopted
parameters, listed in Table \ref{tab:HD166620parameters}, are the weighted average of the results
from these three methods.}
\centering
\begin{tabular}{c c c c}
\hline\hline
Parameter& HD\,166620& HD\,144579  & Method\\ [0.3ex]
\hline 
 & 5005 \textpm 116 & 5323 \textpm 70 & ARES+MOOG\\
$\mathrm{T}_\mathrm{eff}$ (K) & 4966 \textpm 70 & 5248 \textpm 70 & \texttt{CCFPams}\\
\vspace{2mm}
& 4997 \textpm 50 & 5316 \textpm 50 & SPC\\

& 4.82 \textpm 0.22 & 4.84 \textpm 0.12 & ARES+MOOG\\
$\mathrm{log(g) (cgs)}$ & 4.70 \textpm 0.20 & 4.56 \textpm 0.20 & \texttt{CCFPams}\\
\vspace{2mm}
& 4.43 \textpm 0.10 & 4.60 \textpm 0.10 & SPC\\

& -0.21 \textpm 0.06 & -0.63 \textpm 0.05 & ARES+MOOG\\

[Fe/H]  &  -0.25 \textpm 0.05 & -0.70 \textpm 0.05 & \texttt{CCFPams}\\

[m/H]  & -0.19 \textpm 0.08 & -0.62 \textpm 0.08 & SPC\\

\end{tabular}
\label{tab:parametercomparison}
\end{table}
 
\begin{table*}
    
\caption{Stellar parameters obtained for HD\,166620 and HD\,144579 from the literature and from the present study. All methods obtaining the parameters are described in Section \ref{subsec:stellarcharacteristics}} .
\centering
\begin{tabular}{c c c c}
\hline\hline
Parameter& HD\,166620& HD\,144579  & Reference\\ [0.3ex]
\hline 
\vspace{2mm}
RA (J2000) & 18:09:37.4162 & 16:04:56.7936 & {\citet{2021A&A...650C...3G}} \\
\vspace{2mm}
DEC (J2000) & +38 27 27.9980 & +39:09:23.4346 & {\citet{2021A&A...650C...3G}} \\
\vspace{2mm}
$\mu_\alpha$ (mas/yr) & $-316.454\pm0.018$ & $-570.872\pm0.016$ & {\citet{2021A&A...650C...3G}} \\
\vspace{2mm}
$\mu_\delta$ (mas/yr) & $-468.348\pm0.020$ & $52.633\pm0.017$ & {\citet{2021A&A...650C...3G}} \\
\vspace{2mm}
$G$ & $6.130\pm0.003$ & $6.461\pm0.003$ & {\citet{2021A&A...650C...3G}}\\
\vspace{2mm}
B & $7.28\pm0.02$ & $7.388\pm0.065$ & {\citet{{2000A&AS..143....9W}}} \\
\vspace{2mm}
V & $6.39\pm0.02$ & $6.655\pm0.014$ & {\citet{{2000A&AS..143....9W}} } \\
\vspace{2mm}
J & $4.952\pm0.0278$ & $5.182\pm0.02$ &{\citet{{2003tmc..book.....C}} } \\
\vspace{2mm}
H & $4.458\pm0.192$ & $4.824\pm0.017$ & {\citet{{2003tmc..book.....C}} } \\
\vspace{2mm}
K & $4.232\pm0.021$ & $4.755\pm0.016$ & {\citet{{2003tmc..book.....C}} } \\
\vspace{2mm}
W1 & $4.236\pm0.279$ & $4.733\pm0.178$ &{\citet{{2014yCat.2328....0C}} } \\
\vspace{2mm}
W2 & $3.996\pm0.13$ & $4.601\pm0.083$ & {\citet{{2014yCat.2328....0C}} } \\
\vspace{2mm}
W3 & $4.235\pm0.014$ & $4.777\pm0.015$ & {\citet{{2014yCat.2328....0C}} }\\
\vspace{2mm}
Spectral type & K2V & G8V & This work\\
\vspace{2mm}
Parallax (mas)& $90.12 \pm 0.02$ & $69.64 \pm 0.014$ & {\citet{2021A&A...650C...3G}}\\
\vspace{2mm}
 Radial Velocity (km s$^{-1}$) & $-19.51 \pm 0.12$  & $-59.44 \pm 0.12$  &{\citet{2021A&A...650C...3G}}\\
\vspace{2mm}
U (km s$^{-1}$) & $16.80\pm0.05$ & $-35.88\pm0.04$ & This work\\
\vspace{2mm}
V (km s$^{-1}$) & $-31.34\pm0.10$ & $-58.51\pm0.07$ & This work\\
\vspace{2mm}
W (km s$^{-1}$) & $0.25\pm0.05$ & $-18.57\pm0.09$ & This work\\
\vspace{2mm}
T$_\mathrm {eff}$ (K)& $4989\pm48$ & $5296\pm37$& This work$^{a}$\\
\vspace{2mm}
$\log$(g)$_{\mathrm{spec}}$ (cgs) & $4.65\pm0.10$ & $4.67\pm0.08$ & This work$^{a}$\\
\vspace{2mm}
[Fe/H] (dex) & $-0.21\pm0.04$ & $-0.65\pm0.04$ & This work$^{a}$\\
\vspace{2mm}
[$\alpha$/Fe] & $0.20$ & $0.26$ & This work\\
\vspace{2mm}
v$\sin$i (km s$^{-1}$) & $<2.0$ & $<2.0$ & This work\\ 
\vspace{2mm}
$\xi_t$ (km s$^{-1}$) & $0.45\pm0.31$ & $0.62\pm0.12$ & This work\\ 
\vspace{2mm}
 Mass (M$_\odot$) & 0.76$_{-0.019}^{+0.032}$ & 0.73$_{-0.013}^{+0.022}$ &This work\\
\vspace{2mm}
Radius (R$_\odot$)& 0.77$_{-0.006}^{+0.007}$& 0.76$_{-0.004}^{+0.005}$ &This work\\
\vspace{2mm}
$\log$g (cgs)& 4.55$_{-0.01}^{+0.02}$ & 4.53$\pm0.01$ & This work\\
\vspace{2mm}
Density ($\rho_\odot$)& 1.67$_{-0.059}^{+0.093}$ & 1.63$_{-0.038}^{+0.057}$&This work\\
\vspace{2mm}
Luminosity (L$_\odot$)& 0.36$_{-0.01}^{+0.02}$ & 0.47$\pm0.01$ & This work\\
\vspace{2mm}
Distance (pc)& 11.09$\pm0.002$ & 14.36$\pm0.002$ & This work\\
\vspace{2mm}
Age (Gyr)& 10.09$_{-3.76}^{+2.73}$ & 11.90$_{-2.50}^{+1.57}$&This work\\
\vspace{2mm}
\textcolor{black}{log(R' HK) (median)}& \textcolor{black}{-5.143}&\textcolor{black}{-4.976}&\textcolor{black}{This work}\\
 P$_\mathrm {rot}$& 42.4 d &--& {\citet{1981ApJ...250..276V}}\\
 \vspace{2mm}
 P$_\mathrm {cyc}$& 17 years&--&{\citet{2022AJ....163..183B}}\\


\hline
\end{tabular}
\label{tab:HD166620parameters}
\end{table*}

\subsection{Stellar characteristics of HD 166620 and HD 144579}
\label{subsec:stellarcharacteristics}

Galactic velocities for both stars were derived using the Gaia DR3 \citep{2021A&A...650C...3G} data and in particular the trigonometric parallax, the radial velocity, the positions and the proper motions. Following \citet{1987AJ.....93..864J} we calculated the galactic velocities $U$, $V$, and $W$, as reported in Table \ref{tab:HD166620parameters}. These are expressed in the directions of the Galactic center, Galactic rotation, and north Galactic pole, respectively, and we did not subtract the solar motion from our calculations. Using this kinematical information, we can estimate which population in the Galaxy our stars belong to. Using the descriptions in \citet{2006MNRAS.367.1329R}, we find that there is $98.5\%$ and $90.8\%$ probability that HD\,166620 and HD\,1445797, respectively, belong to the thin disc. 

We utilised our HARPS-N spectra (details below) to estimate the stellar atmospheric parameters: effective temperature (T$_\mathrm {eff}$), metallicity [Fe/H], surface gravity, microturbulent velocity and projected rotational velocity. Following \citet{2020MNRAS.499.5004M}, three independent methods were used for obtaining these parameters: (1) ARES+MOOG\footnote{ARESv2: \url{http://www.astro.up.pt/~sousasag/ares/}; MOOG 2017: \url{http://www.as.utexas.edu/~chris/moog.html}}, using a stacked spectrum for each star, where the stitched spectra were simply added together after putting them in the lab-frame. Parameters were estimated using a curve-of-growth method based on neutral and ionised iron lines \citep{2014dapb.book..297S}. (2) \texttt{CCFPams}\footnote{\url{https://github.com/LucaMalavolta/CCFpams}}, using the Cross Correlation Functions and applied an empirical relation, found in \citet{2017MNRAS.469.3965M}. (3) Finally, spectral synthesis was applied on all individual spectra using the Stellar Parameter Classification tool \citep[SPC - ][]{2012Natur.486..375B}. 

Surface gravities are notoriously hard to measure accurately from spectra. The values resulting from ARES+MOOG and \texttt{CCFPams} were corrected following \citet{2014A&A...572A..95M} while the value for SPC was additionally constrained using YY isochrones \textcolor{black}{\citep{2001ApJS..136..417Y}}. Precision errors for effective temperature, surface gravity and metallicity from ARES+MOOG and \texttt{CCFPams} were inflated for accuracy by 60\,K, $0.1$\,dex and $0.04$\,dex respectively following \citet{2011A&A...526A..99S}. Errors for the values of SPC are also accounted for accuracy \citep{2012Natur.486..375B}. We note that ARES+MOOG is the only method to derive microturbulent velocity ($\xi_t$) while SPC is the only method that derived the projected rotational velocity ($v\sin i$). 


The stacked spectra were additionally used to derive individual chemical abundances for magnesium, silicon and titanium. To calculate these abundances, ARES+MOOG was used with more details provided in \citet{2013A&A...558A.106M}. These three abundances, in combination with the iron abundance, were then used to calculate the alpha-over-iron abundance, $[\alpha/$Fe$]$. We find that HD\,166620 has an alpha element overabundance of $0.20$ and HD\,144579 of $0.26$, calculated with respect to iron as compared to the Sun. That makes these stars as alpha-enhanced as the planet hosts K2-111 \citep{2020MNRAS.499.5004M} or TOI-561 \citep{2021MNRAS.501.4148L} whose small planets have densities that are significantly below the densities of small planets around non-alpha-enhanced stars.

Stellar mass, radius, density, luminosity, distance, and age were further obtained from an isochrones and evolutionary tracks analysis, as described in \citet{2020MNRAS.499.5004M}. We used the three individual values for effective temperature and metallicity, the Gaia DR3 parallax, apparent magnitude in 8 photometric filters (listed in Table \ref{tab:HD166620parameters}), and two different sets of isochrones (Dartmouth and MIST). For each parameter, six sets of posterior distributions were obtained using the code \texttt{isochrones} \citep{2015ascl.soft03010M} and the nested sampling algorithm \texttt{MultiNest} \citep{2019OJAp....2E..10F}. All six scenarios yielded consistent results. The final adopted parameters, as listed in Table \ref{tab:HD166620parameters}, were calculated from the median and 16th and 84th percentile of the merged posterior distributions. A more precise value for surface gravity was then calculated using the stellar mass and radius.

\begin{table*}
\caption{Prior distributions used in the radial velocity model in the {\sc kima}  analysis} 
\centering
\begin{tabular}{c r r r r r}
\hline\hline
Parameter& Notation&Unit & HD\,166620& HD\,144579 & Distribution \\ [0.2ex]
\hline 
Number of Keplerians to fit&$N_{\rm p}$  &         & $\cal{U}$(0, 5)       & $\cal{U}$(0, 5)& Uniform\\
Orbital period&$P$    &days     &$\cal{U}$(1.1, 3492)   & $\cal{U}$(1.1, 2700)& Uniform\\
RV Semi-amplitude&$K$    & m s$^{-1}$&$\cal{MLU}$(0.2, 15)  &$\cal{MLU}$(0.2, 4)& Modified Log-Uniform\\
Eccentricity&$e$    &         &$\cal{K}$(0.867, 3.03) &$\cal{K}$(0.867, 3.03)& Kumaraswamy\\
Orbital phase&$\phi$ &         &$\cal{U}$(0, 2$\pi$)   &$\cal{U}$(0, 2$\pi$)& Uniform\\
Longitude of line of sight&$\omega$&         &$\cal{U}$(0, 2$\pi$)  &$\cal{U}$(0, 2$\pi$)& Uniform\\
Decorrelation vectors&$\beta$ &m s$^{-1}$& $\cal{G}$(0, 10)   & $\cal{G}$(0, 10)& Gaussian \\
Systemic velocity& $\gamma$ & m s$^{-1}$ & $\cal{U}$(-2.89,5.37) & $\cal{U}$(-4.84,4.96) & Uniform\\
White noise&$\sigma_{\rm jit}$& m s$^{-1}$&$\cal{LU}$(0.01, 10 x rms)&$\cal{LU}$(0.01, 10 x rms)&Log Uniform

 \\[1ex] 
\hline
\end{tabular}
\label{tab:priors}
\end{table*}


\begin{figure}
    \centering
	
    \includegraphics[width=\columnwidth]{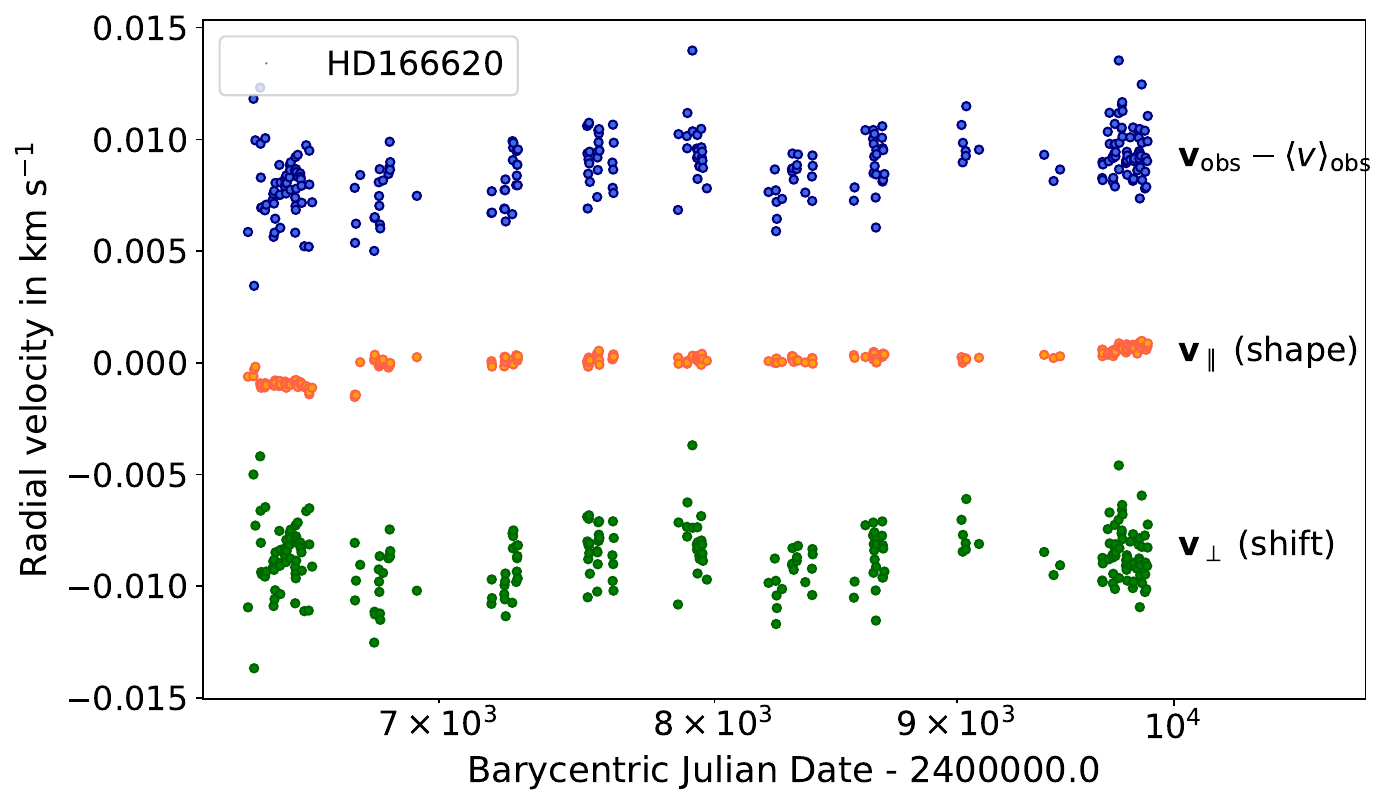}
    \includegraphics[width=\columnwidth]{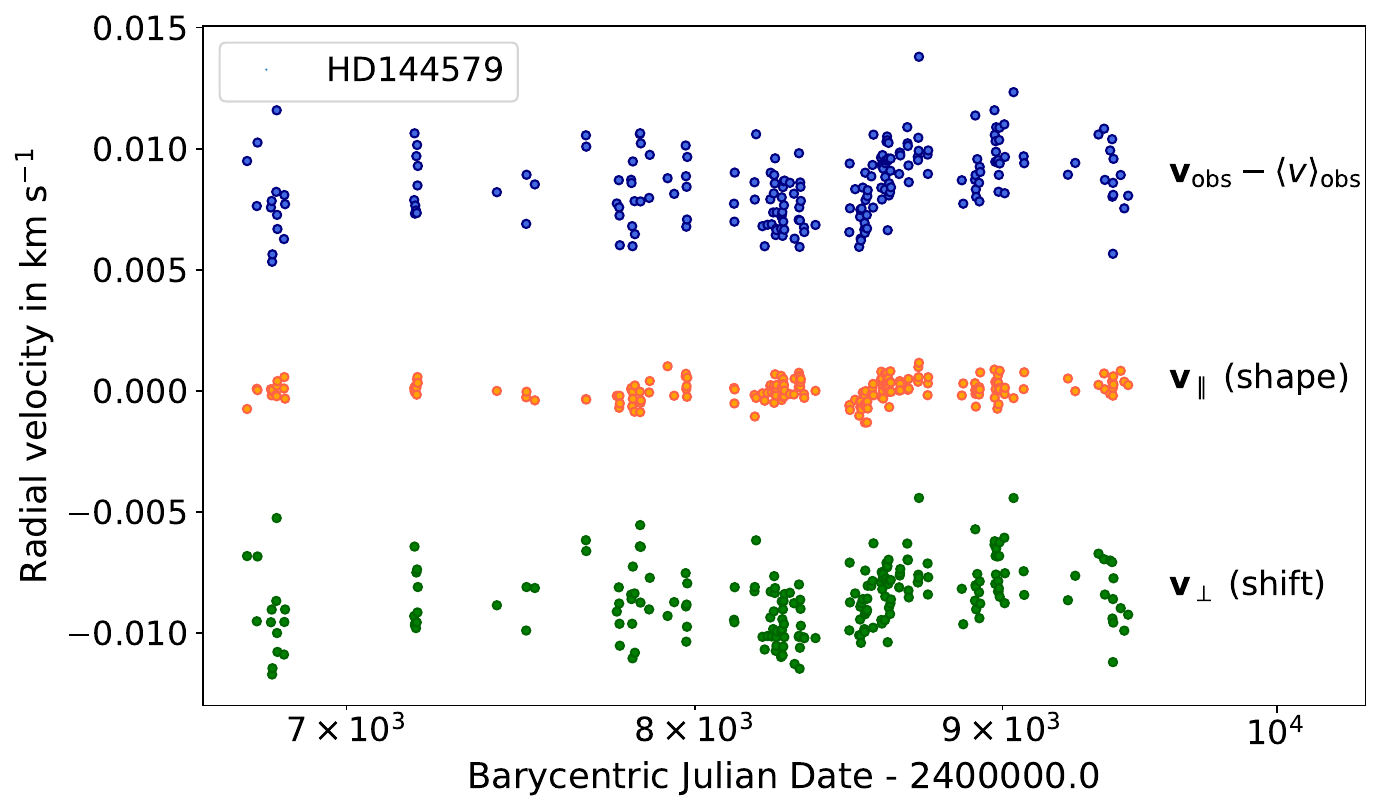}
	
    \caption{  In both panels, the blue scatterplot at the top is the barycentric RV with its own mean subtracted. The orange time series in the middle represents  the shape-driven component obtained from the {\sc scalpels} projection, while the green one shows the `cleaned' shift-driven velocities, obtained when the shape-driven variations are subtracted from the observed RVs. Please note that an offset is introduced to each RV component for better illustration. \textcolor{black}{The jump in the shape time series of HD166620, at around BJD= 2456737 corresponds to the focus intervention that occurred in the HARPS-N instrument.}}
    \label{fig:scalpels}
\end{figure}

\section{Analyses and results}
\subsection{RV analysis pipeline}
\label{subsec:pipeline}
We employed {\sc tweaks} as an RV analysis pipeline to search for planetary signals below the 1 m s$^{-1}$ RV barrier imposed by stellar activity (Figure\,\ref{fig:submsbarrier}).
This pipeline uses the {\sc scalpels} technique to separate the stellar variability component in the RVs, driven by spectral line-shape changes \citep{2021MNRAS.505.1699C}. The temporal variability of the CCF contains both spectral line `shape' changes and Doppler `shifts'. {\sc scalpels} uses the translational invariance property of the autocorrelation function (ACF; \citet{adler1962}) of the CCF to isolate the effects of shape changes in the CCF from shifts of dynamical origin. {\sc scalpels} constructs an orthogonal basis containing the coefficients of the first few principal components of shape-induced CCF variations, then projects the raw RVs onto this basis to obtain a time series of shape-induced RV variations. These are then subtracted from the original RVs to leave shift-only RV variations, where the planet signals are sought \citep{2022MNRAS.515.3975A}.

The perturbations resulting from stellar activity (shape component, shown as the orange time-series in Figure\,\ref{fig:scalpels}) are isolated by  projecting the RVs onto the time-domain subspace spanned by the amplitude coefficients (a.k.a $\mathbf{U}$-vectors)  of the ACF's principal components \citep{2021MNRAS.505.1699C}. However, when projected onto the orthogonal complement of the time-domain ($\mathbf{U}$) subspace, the dynamical shifts (shift component, shown as the green timeseries in Figure\,\ref{fig:scalpels}) due to the planets are preserved. 

\textcolor{black}{We found that the first three leading principal components are the strong contributors to the shape component out of the first five (see Figure\,\ref{fig:HD166620corner} \& \ref{fig:HD144579corner}). This optimal ranking was done using leave-one-out cross-validation (LOOCV). \citet{2021MNRAS.505.1699C} found this method
to be efficient in identifying the number of leading columns
of \textbf{U} that contain significant profile information. The raw implementation of {\sc scalpels} returns not just the basis vectors, but also the RV response. However, the shape time series in Figure\,\ref{fig:scalpels} are from the {\sc scalpels}-only implementation, not the coefficients of the amplitude obtained from the simultaneous fitting of any keplerian.}

The basis vectors ($\mathbf{U}$-vectors) representing the shape components identified by {\sc scalpels} are then used for stellar activity decorrelation using the {\sc kima} nested sampling package  \citep{2018JOSS....3..487F}. {\sc kima} employs Diffusive Nested Sampling (DNS \citet{2014arXiv1411.3921B}) to sample the posterior distributions for each of the orbital parameters by modelling the RV data with a sum of up to N$_{\rm p}$ Keplerian functions. This gives {\sc kima}  the estimates for the Bayesian evidence ($\cal{Z}$) of the data supporting each model, enabling the model comparison \citep{2010ascl.soft10029B}. The ratio of Bayesian evidences between models with different N$_{\rm p}$ can then be compared. There is a `known' object setting provided by {\sc kima}, which allows one set of priors to be applied to the orbital parameters of a known planet while a different set of priors is used to explore the existence and parameters of putative planets \citep{2022MNRAS.511.3571S}. When a candidate Keplerian signal is proposed in the model, {\sc kima} conducts fits for both the `known' planet and the new signal. This option is useful if there is a transiting planet with precisely determined transit parameters or another `known' Keplerian signal in the system that we can confidently fit out. \citet{2022MNRAS.515.3975A}, presented a pilot  demonstration of {\sc tweaks} in action. They used a GP along with {\sc scalpels} + {\sc kima}  to update the transiting planet's mass and establish the existence of a third planet in the CoRoT-7 system.

 {\sc kima} posteriors are also used to estimate the detection limits as in \citet[e.g.,][]{2022MNRAS.511.3571S,2023arXiv230110794S} and  \citet{2022MNRAS.514.2259S} for the entire orbital period range spanned by the data. With {\sc kima} , we test our sensitivity to detecting low-mass planets and use it to generate a reliable limit of detection before and after correcting for the stellar variability using {\sc scalpels}. Additionally, following \citet[][]{2022MNRAS.511.3571S,2023arXiv230110794S} we evaluate the effectiveness of our detection protocol by injecting low-mass planets with RV semi-amplitudes in the sub-m\,s$^{-1}$ level into our HARPS-N data and retrieving them with {\sc kima} and {\sc scalpels}.

\subsubsection{Prior distribution}

The models we fit for the two stars are defined by the priors given in Table \ref{tab:priors}. These priors are comparable to those described in \citet{2020A&A...635A..13F}, but tailored for HD\,166620 and HD\,144579 independently.

We set the number of Keplerians $N_{\rm p}$ to be a free parameter with a uniform prior between 0 and 5. 
i.e, {\sc kima}  will try to fit a maximum of 5 Keplerians to the data simultaneously and independently of each other. For this search, we use a uniform distribution for $\phi$, $\gamma$ and $\omega$ , as there is no justification to favour any specific value within these parameter regimes. We use a Kumaraswamy distribution ($\alpha$= 0.867 and $\beta$= 3.03) \citep{1980JHyd...46...79K} for planetary eccentricities, which favours lower values but still permits the exploration of higher eccentricities when the data requires it \citep{2022MNRAS.511.3571S}. The values for shape parameters $\alpha$ and $\beta$ are justified by the Beta distribution in \citet{2013MNRAS.434L..51K}.

\begin{figure}
    \centering
    \includegraphics[width=\columnwidth]{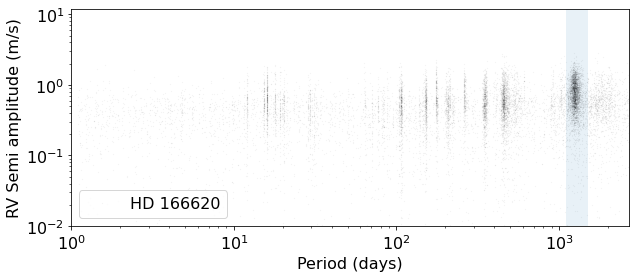}
	\includegraphics[width=\columnwidth]{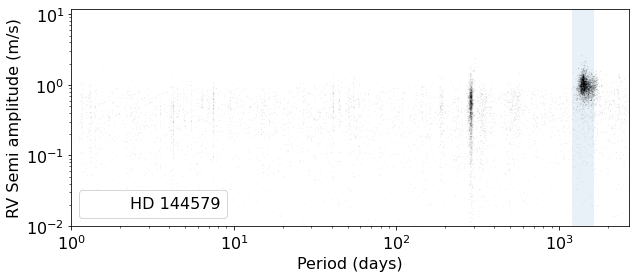}
    \caption{The {\sc kima} joint posterior distribution of RV semi-amplitudes (in log scale) for HD\,166620 and HD\,144579 in log orbital period space are shown. 
    Both stars exhibit an anomalous signal with a similar orbital period of 1400 days. The posterior orbital period space is populated with more over-density patches, which represent signs of additional signals, possibly planets, aliases and/or harmonics. 
    }
    \label{fig:1500days}
\end{figure}

The stellar reflex orbital semi-amplitude ($K$) is the one parameter to sample with the highest sensitivity when establishing robust detection limits \citep{2022MNRAS.511.3571S}. We will describe this analysis in Section \ref{subsec:detectionlimits}. In this regard, we use a modified Log-Uniform (Jeffreys) prior. 
The prior values given for $K$ in Table \ref{tab:priors} are obtained
by carrying out a number of initial runs with uninformative priors, to establish a plausible upper limit.
The model also accounted for the systemic velocity of the centre of mass of the system ($\gamma$: which corresponds to an RV offset measured by HARPS-N) and a jitter term ($\sigma_{\rm jit}$) added in quadrature, to effectively represent the uncorrelated noise sources (uncorrelated at the timescale of the planet signals we are looking for) such as granulation and small night-to-night wavelength calibration errors.
 For $\sigma_{\rm jit}$, we  again use a Jeffreys prior. In order to account for the stellar activity, the {\sc scalpels} U-vectors, representing the projection of line-shape changes onto the RV are included as decorrelation vectors (as in \citet{2022MNRAS.515.3975A}), using a Gaussian prior for their amplitude coefficients.

\subsection{1400 day periodicity}
\label{sub:instrumental}

\begin{figure}
    \centering
	\includegraphics[width=1\columnwidth]{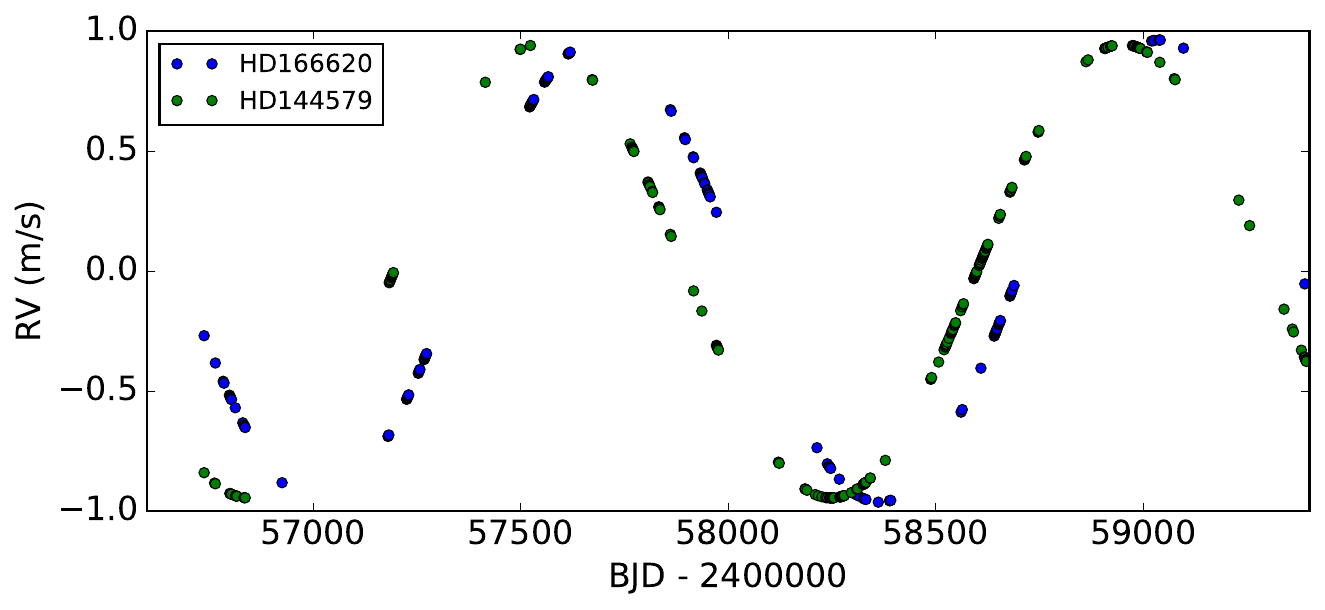}

    \caption{ The orbits have been simulated using the best-fit parameters for the $\sim$ 1400d signal, observed in HD\,166620 and HD\,144579. The signals are nearly in phase (there is a small offset) and exhibit comparable RV variations of 95 cm s$^{-1}$. The phase offset possibly arises from the slightly different periods (1390 and 1420 days) of these signals.}
    \label{fig:phaseplot}
\end{figure}

 Figure\,\ref{fig:1500days} shows the resultant posterior distribution for the RV semi-amplitudes for HD\,166620 and HD\,144759, plotted against the posterior orbital periods. The analyses initially revealed a long-term periodicity of approximately 1400 days in both stars. We obtained meaningful fits for 1400-day sinusoids in both stars while decorrelating against the {\sc scalpels} shape-driven variations. We found that these signals are also more or less in phase and share very similar sub-m s$^{-1}$ (95 cm s$^{-1}$) semi-amplitudes (Figure\,\ref{fig:phaseplot}). This suggested a systematic, rather than a dynamical origin. Therefore, we investigated further the nature and possible origin of these common periodic signals. 

 \subsubsection{Instrumental zero-point subtraction}
\label{subsec:zeropoint}
We searched for a common signal with a similar orbital period ($\sim$ 1400 d), phase, and semi-amplitude in 12 of the most intensively-observed RPS targets to gain clarity on whether or not this signal is common to all the RPS targets. From parallel analyses using {\sc yarara} \citep{2021A&A...653A..43C} and {\sc scalpels} \citep{2021MNRAS.505.1699C}, we discovered sporadic sub-m s$^{-1}$ jumps in the observed RV data of most of the RPS targets, in a pattern that mimics a long-term periodicity. Further investigation revealed that these discontinuities in RV occurred at the times of maintenance interventions performed on the HARPS-N instrument.  
These were the periodic warm-ups performed on the detector as a temporary solution for the small leak in the detector cryostat \citep{2021A&A...648A.103D}. The detector cryostat was changed in October 2021. This solved the leak problem, but the jumps remain in observations made prior to this intervention.


\begin{figure}
    \centering
	\includegraphics[width=1\columnwidth]{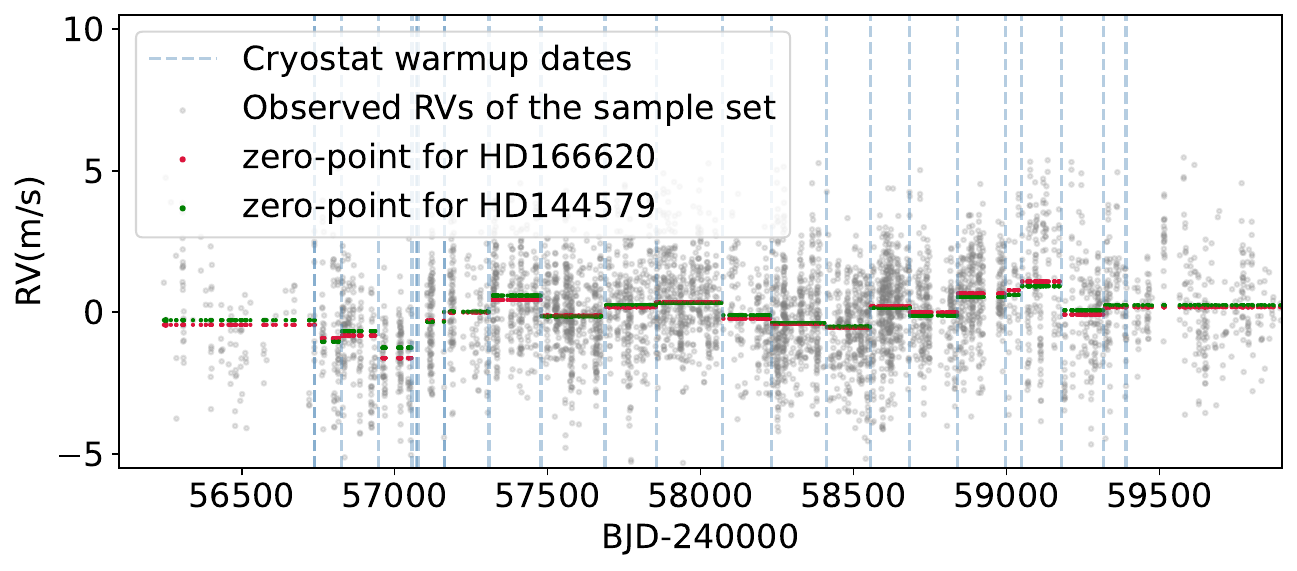}
    \caption{The estimated zero points in RVs between the epochs of cryostat warmups for HD\,166620 and HD\,144579 are shown by red and green points respectively. It is evident that when we analyse HD\,166620 and HD\,144579 individually, we find very similar values for the offsets.
    The grey points in the background represent the median-subtracted shift RVs of 11 other HARPS-N targets, corrected for activity variations using {\sc scalpels}. The dashed grey vertical lines indicate the dates of instrumental interventions considered for the zero point estimations. }
    \label{fig:zeropoints}
\end{figure}

We experimented with establishing a zero-point offset between the interventions and subtracting it from the target star RVs in order to account for these substantial instrumental interventions (cryostat warmups). A sample set of stars with long observation baselines, good seasonal sampling, and photon uncertainties of less than 1 m s$^{-1}$ was selected for this purpose. Table \ref{tab:offsetsample} displays the list of stars (including the Sun) chosen using these criteria, the majority of which shared a comparable long-term periodicity. This enabled better quantification of the RV offsets between major interventions performed  on the instrument.To prevent removing genuine signals from the data of HD\,166620 and HD\,144579 analysed in this paper, we removed those stars from the sample used to calculate the zero point offsets induced by detector warmups. i.e, we omitted HD\,144579 from the sample set to calculate the zero points for HD\,144579 and did the same for HD\,166620.

The zero-point estimation was performed as follows. The observed RVs of each star were first corrected for activity cycles by removing the {\sc scalpels} shape component. We then removed the extreme outliers after subtracting the median RV from the activity-corrected shift RVs. 
We divided the data set into segments between the known dates of interventions and estimated the zero point as the median of the combined corrected RV data set in each segment. The floating chunk offset (FCO) technique, used to correct for night-to-night zero-point velocities by \citet{2014A&A...568A..84H, 2011ApJ...743...75H, 2010A&A...520A..93H}, served as the basis for chunk-by-chunk offset subtraction used in this study. The dates of the cryostat warmups that define the segment boundaries while estimating the zero points are represented by the vertical lines in Figure\,\ref{fig:zeropoints}. Then, we simply subtract the estimated chunk-by-chunk zero-point (illustrated in red and green in Figure\,\ref{fig:zeropoints}) from the observed RV data of the target stars for further investigation. \textcolor{black}{The hence obtained BJD boundaries and zero-point offsets are given in Table\ref{tab:RVoffsets}.} This finding serves as a powerful demonstration of the ability of HARPS-N to monitor such a sub-m s$^{-1}$ level signal over a ten-year period.

\begin{figure}
    \centering
	
    \includegraphics[width=\columnwidth]{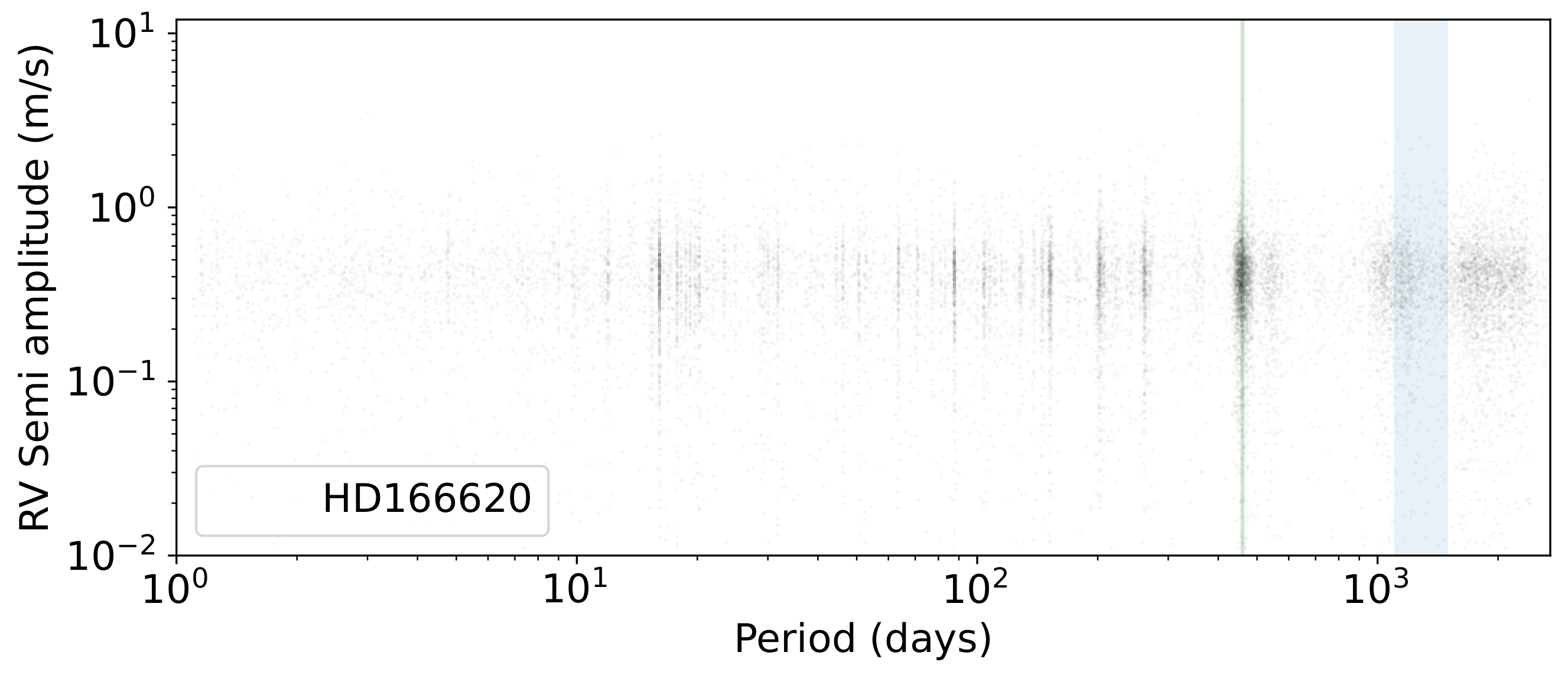}
    \includegraphics[width=\columnwidth]{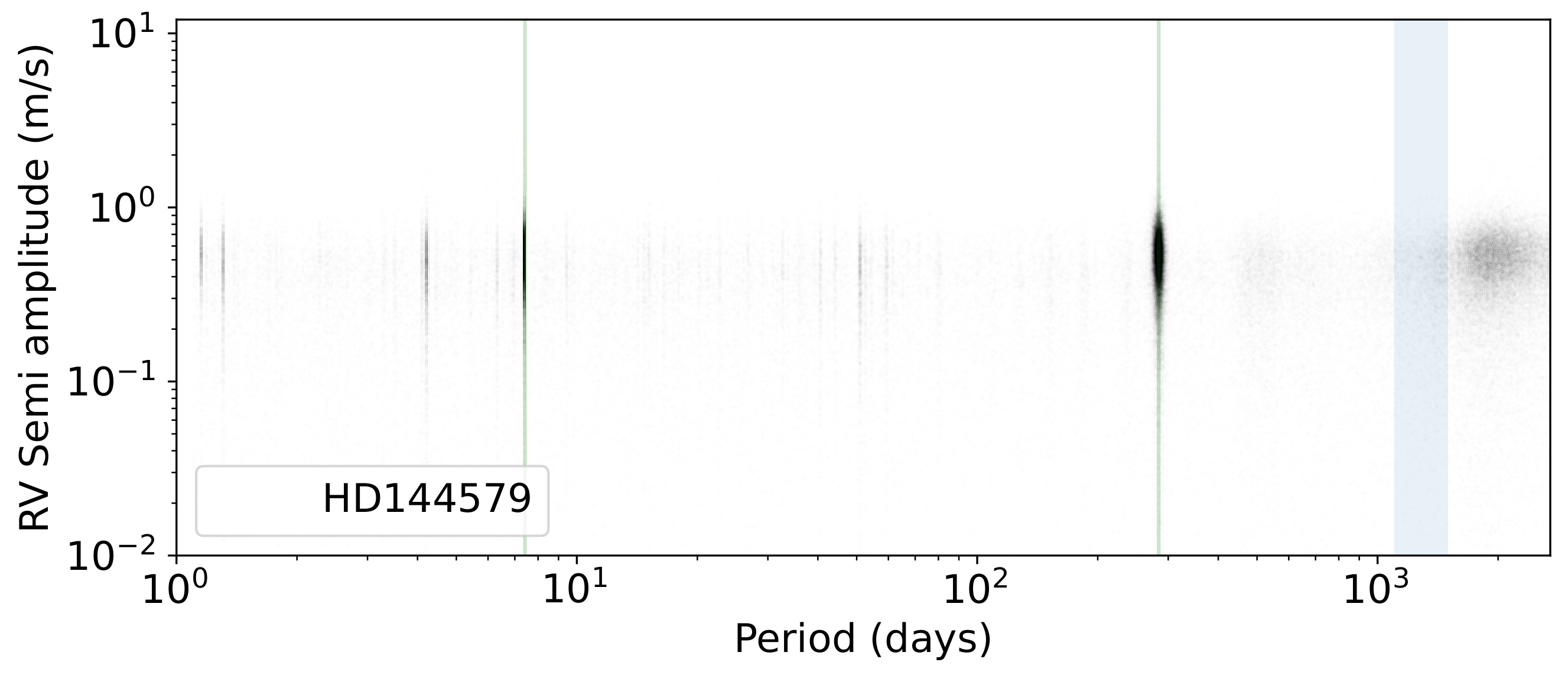}
	
    \caption{ The joint posterior distributions from {\sc kima} after subtracting the  zero points estimated for the chunks of observations between major cryostat interventions, shown in Figure\,\ref{fig:zeropoints}. The common long-term instrumental signal shown in Figure\,\ref{fig:1500days} with orbital period $\sim$1400 d is hence modelled out. The displayed posteriors are obtained from more than 100000 independent trial models for HD\,166620 (Top) and HD\,144579 (Bottom).}
    \label{fig:raw-offset}
\end{figure}

\subsection{Posterior analysis}
\label{posteriors}
\subsubsection{HD\,166620: The tale of a  `quiet' star}

Figure\,\ref{fig:raw-offset} displays the RV semi-amplitude posteriors from {\sc kima}, after they have been adjusted for the long-term instrumental systematics (using the zero-point subtraction method outlined in \ref{subsec:zeropoint}). The prevalent 1400 d signal that was caused by occasional cryostat warmups is no longer dominant (compare Figures \ref{fig:1500days} and \ref{fig:raw-offset}). 

We then directed our search for planet candidates in the offset-subtracted RVs using {\sc tweaks} (described in section \ref{subsec:pipeline}) that takes into consideration the instrumental shifts and stellar activity cycles in both the wavelength and time domains. Using the {\sc kima}  nested-sampling package \citep{2018JOSS....3..487F} and a model with up to five unknown Keplerian signals, we performed a blind search of the radial velocities. With a sum of Keplerian functions from N$_{\rm p}$ orbiting planets, the algorithm models the RV time series while estimating the posterior distributions for all of the orbital parameters. With {\sc scalpels} \citep{2021MNRAS.505.1699C}, which uses principal-component analysis of the autocorrelation function of the CCF, the time-domain activity-decorrelation vectors were computed. These vectors are then used as independent activity indicators for linear decorrelation in {\sc kima} . 

The top panel of Figure\,\ref{fig:raw-offset} displays the resultant joint posteriors for HD\,166620 after the removal of 
the floating-chunk zero-point estimates between cryostat warm-ups
and decorrelation against the {\sc scalpels} decorrelation vectors. The posterior samples below the detection threshold appear as a background `fog' following the priors for $K$ and log P.

A relatively well-defined signal appears at an orbital period of 460.59 d. Using a Gaussian mixture model based on \citet{2014zndo.15856}, 
we computed the probabilities of posterior samples belonging to 
Gaussian foreground and background populations in $\log K$ and $\log P$, to exclude the posterior points coming from the uniform prior `fog'. This way, we isolated the foreground probability of the cluster and obtained a reliable estimation for the RV semi-amplitude for this signal as 
$0.41\pm0.10$
m s$^{-1}$. 
If this signal is planetary, 
this would correspond to a planet with an upper $M\sin i$ mass limit of
$3.34\pm 0.85$
\mearth, when taking into account a stellar mass of 0.76 \msun. This method increases the precision and accuracy of $K$ compared to the standard measurement of the posterior in {\sc kima}.

We conducted a False Inclusion Probability (FIP) analysis \citep{2022A&A...663A..14H} in frequency space, with the bin size set to the Nyquist frequency resolution over the entire data duration. 
The Gaussian mixture model allowed us to calculate the probability of posteriors in the foreground, independently of the choice of frequency bin width. We first computed the True Inclusion Probability (TIP), as the fraction of all {\sc kima} 
trial models
(N$_{\rm s}$) that 
contain planets whose periods
fall in the frequency interval of interest, belonging to the foreground. In other words, TIP is the number of models for which the orbital period (P) and RV semi-amplitude (K) of a sampled planet fall within their respective foreground populations.
\begin{equation}
    \mathbf{TIP} = \frac{\sum_{i=1}^{\rm N_s}{\rm Pr_i}(\rm foreground)}{\rm N_s}
\end{equation}
The FIP was then obtained as:
\begin{equation}
    \mathbf{FIP} = 1-\mathbf{TIP}
\end{equation}

\begin{figure}
    \centering
	
    \includegraphics[width=\columnwidth]{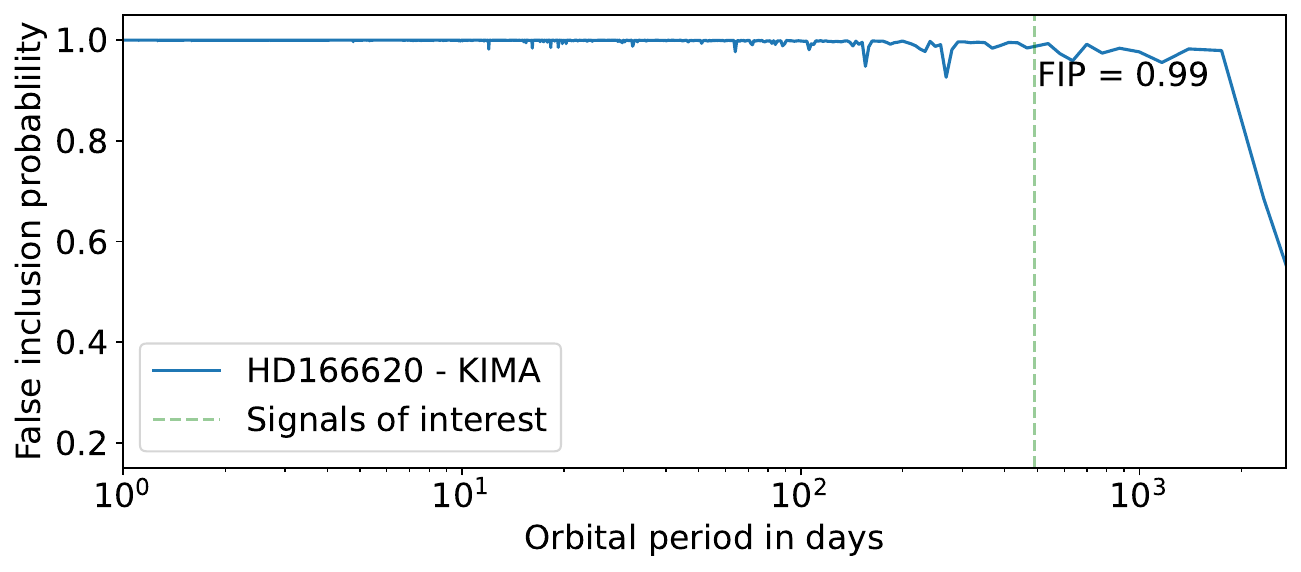}
    \includegraphics[width=\columnwidth]{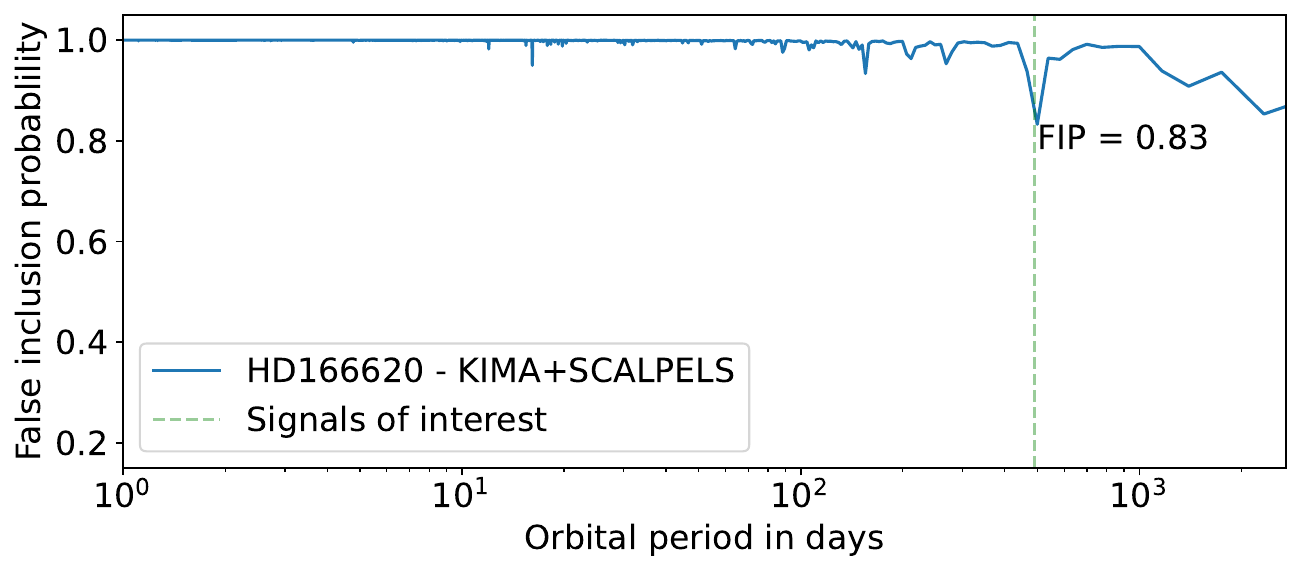}
	
    \caption{ False Inclusion Probability (FIP) periodogram of HD\,166620 showing the minimal values at orbital periods of potential detections. Top : FIP when no {\sc scalpels} U-vectors are used for activity decorrelation in {\sc kima}. Bottom: A more significant FIP value is observed at the same orbital periods when {\sc scalpels} is included. }
    \label{fig:raw-offset_FIP_HD166620}
\end{figure}

Hence, we simultaneously searched for multiple planets using a frequency window sliding across the full frequency range covered by the posterior distribution.  A 
FIP value of 0.83 (Figure\,\ref{fig:raw-offset_FIP_HD166620}) was found at an orbital period of 460.59 d. 
This is not what we anticipate from a strong planetary signal detection, particularly for a star that is magnetically quiet. 
A FIP value of 0.83 indicates that only 17\% of
the models tested favoured a detection at this orbital period interval. To validate this, we performed several analyses by shuffling 
the observing seasons between years and found that the long-term signal at  460.59 d is 
comparable in its amplitude and its FIP to window-function artefacts appearing at other periods in the permuted datasets. We conclude that it too is
most likely to be a window-function artefact.

\subsubsection{HD\,144579: A moderately active `lone' star}
\label{subsec:HD144579}

\begin{figure}
    \centering
	
    \includegraphics[width=1\columnwidth]{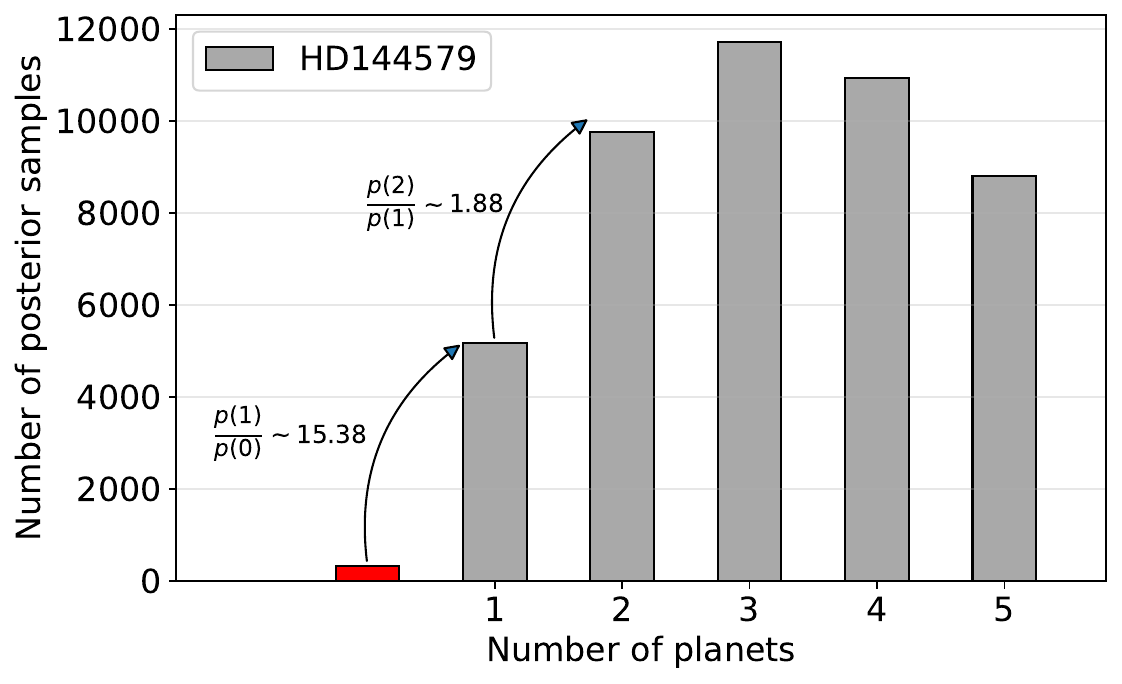}
    
    \caption{The posterior distribution for the number of planets N$_{\rm p}$. The counts are the number of posterior samples in trial models with a specific number of planets. The probability ratios between models with 0,1 and 2 planets are shown. The posterior distribution suggests (Figure\,\ref{fig:raw-offset}) that there are likely to be planets present. 
    The `confident detection' criterion adapted by {\sc kima}, however, takes into account the ratio of probabilities for successive values of N$_{\rm p}$, which favours N$_{\rm p}$=0, rather than depending solely on the probability values. 
     \textcolor{black}{The preferred model (N$_{\rm p}$=0) appears underrepresented, because of the significantly larger parameter space for models with successively greater N$_{\rm p}$.} On the other hand, when a Keplerian was injected, {\sc kima} favoured N$_{\rm p}$=1 (Figure\,\ref{fig:raw-offset_injected80cmhist}).
    \textcolor{black}{A similar representation for HD\,166620 is shown in Fig:\ref{fig:raw-offset_HD166620hist}.}}
    \label{fig:nphist}
\end{figure}

Unlike in HD\,166620, the joint posteriors from the HD\,144579 RVs adjusted for long-term systematics show strong signals at 7.39 and 284.13\,d (bottom panel of Figure\,\ref{fig:raw-offset} ) with RV semi-amplitudes of $0.50 \pm 0.11$ and $0.60 \pm 0.07$ m s$^{-1}$. If these signals are
planetary, they would correspond to planets with
$M\sin i$ of $1.28\pm0.29$ \mearth and $4.14\pm 0.52$ \mearth, respectively, from the foreground posterior probability contained using a Gaussian mixture model. The obtained results favoured $\geq$4.5$\sigma$ detections for both signals. However, based on the ratio of probabilities of consecutive values of the number of planets iterated in the {\sc kima} model comparison, the sampler favoured zero planets (based on the 150 Bayes Factor threshold), as shown in Figure\,\ref{fig:nphist}.
\begin{figure}
    \centering
	
    \includegraphics[width=\columnwidth]{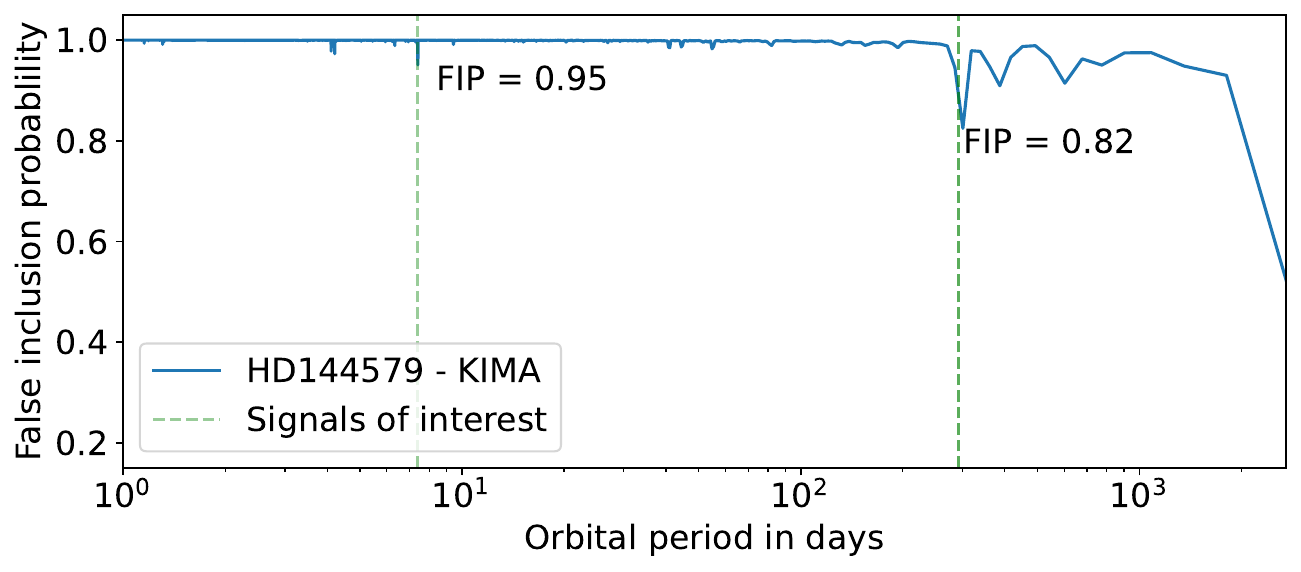}
    \includegraphics[width=\columnwidth]{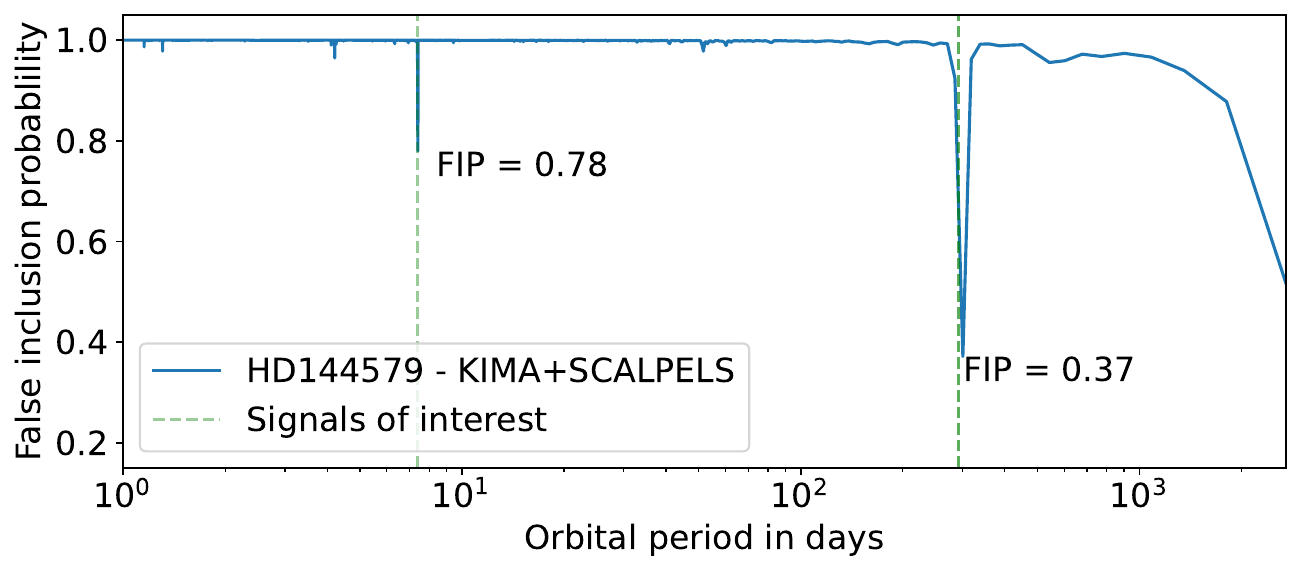}
	
    \caption{ False Inclusion Probability (FIP) for HD\,144579 before and after stellar activity decorrelation. The top and bottom panels are as in Figure\,\ref{fig:raw-offset_FIP_HD166620}. Here, the FIP periodogram of HD\,144579 shows clearer detection of the signals of interest, when decorrelated against the {\sc scalpels} U-vectors. }
    \label{fig:raw-offset_FIP_HD144579}
\end{figure}

\begin{figure}
    \centering
	
    \includegraphics[width=\columnwidth]{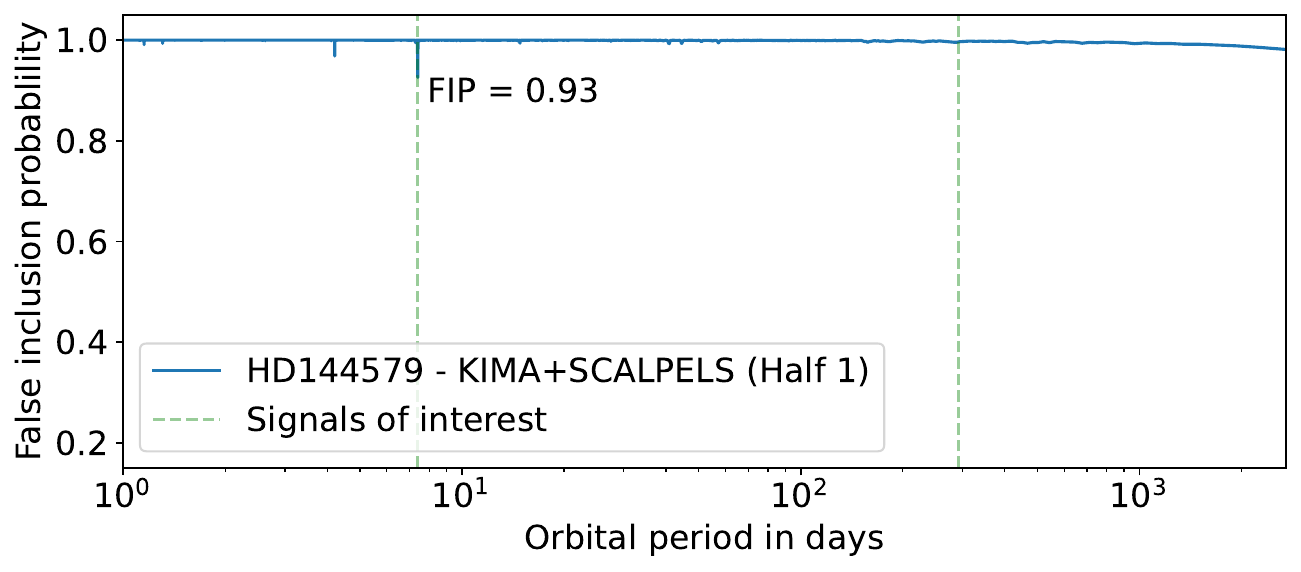}
    \includegraphics[width=\columnwidth]{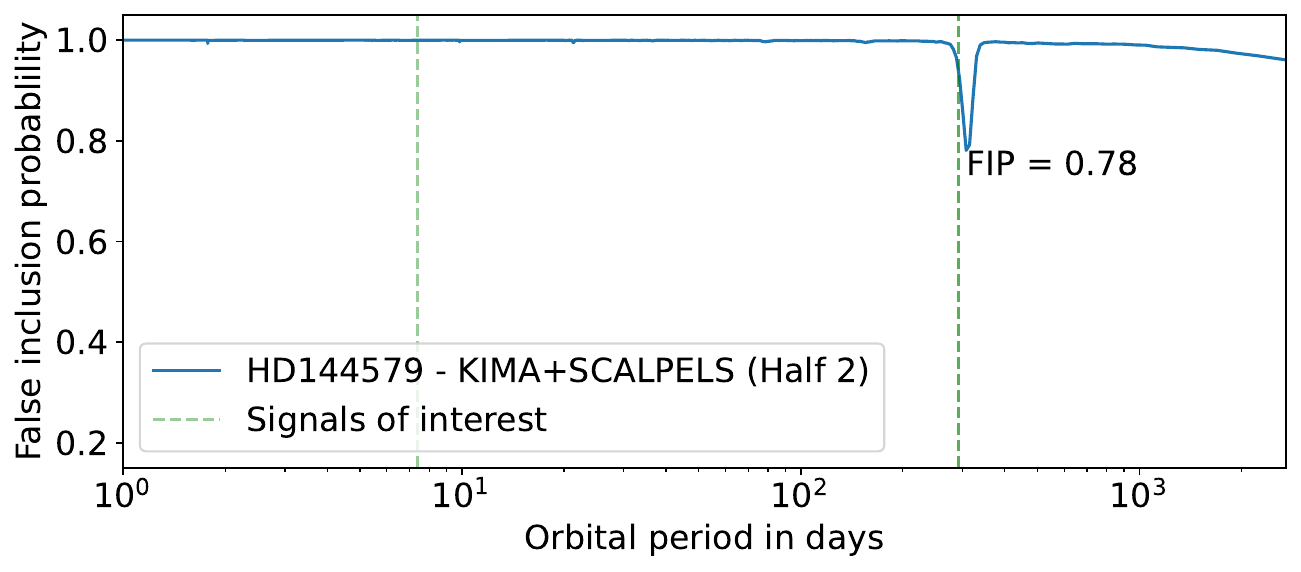}
	
    \caption{ False Inclusion Probability (FIP) for HD\,144579 before and after stellar activity decorrelation. The observations are now divided into two subsets to investigate the coherency of the signals of interest that appeared in the bottom panel of Figure\,\ref{fig:raw-offset_FIP_HD144579}. The FIP diagrams obtained for individual half subsets of the HD\,144579 data are shown in the two panels. Each data subset covered an observation baseline of $>$ 1500 days. Unlike any coherent signal, none of the signals of our interest was commonly detected in both data subsets. Hence, each of these signals could possibly be an artefact of the sampling pattern arising from individual seasons. }
    \label{fig:raw-offset_2halves_HD144579}
\end{figure}

\begin{table}
\begin{footnotesize}
\caption{Derived posterior values for Signals of Interest (SOIs) from the {\sc kima}+{\sc scalpels} RV analysis detailed in Section \ref{posteriors}. The FIP values given are obtained from the entire data. $M_\mathrm{p}$sin$i$ values are estimated considering if the SOIs are planetary.}
\vspace{0.1cm}
\begin{center}
\label{tab:planet_posteriors}
\begin{tabular}{c c  c c }
\hline\hline                
 Parameter (unit) & HD\,166620\, & HD\,144579\, & HD\,144579\,   \\ 
 &SOI.1&SOI.1&SOI.2\\
\hline 
\multicolumn{4}{c}{\textit{Derived parameters from {\sc kima}  posteriors }} \\ 
\hline 
\vspace{0.1cm}
$P$ (d) & 459.88 \textpm 11.05 & 7.39 \textpm 0.002 & 284.13 \textpm 4.16  \\ 
\vspace{0.1cm}
$K$ (m s$^{-1}$)& 0.41 \textpm 0.10 & 0.50 \textpm 0.11 & 0.60 \textpm 0.07 \\ 
$M_\mathrm{p}$sin$i$ (\mearth)& 3.34 \textpm 0.85 & 1.28 \textpm 0.29 & 4.14 \textpm 0.52\\
$e$ & 0.22 & 0.14  & 0.18\\ 
$\omega$ (deg) & 2.97 &  3.48 & 3.01\\ 
$T_0$ (BJD-2400000)& 57491.32641 & 58514.54291  & 58243.86417\\ 
FIP & 0.83 & 0.78 & 0.37 \\ 
\hline 

\end{tabular}
\end{center}
\end{footnotesize}
\label{tab:results}
\end{table}

The False Inclusion Probabilities from the posteriors of HD\,144579 are shown in  Figure\,\ref{fig:raw-offset_FIP_HD144579}.  In contrast to HD\,166620, 
substantially lower FIPs are observed in HD\,144579 at orbital periods of 7.39 and 284.13 d. Although there is a notable 
decrease in the FIP when {\sc scalpels} decorrelation vectors are included in the model, the values are still not 
conclusive.
Using the 284.13 d signal as an example, reporting a detection based on a FIP value of 0.37 (Figure\,\ref{fig:raw-offset_FIP_HD144579}) is 
analogous to boarding an aeroplane with only a 63\% chance of making it to the destination.

Moreover, a number of ambiguous detections can be triggered at spurious periods because of the sampling pattern and cross-talk between various aliases \citep[e.g.,][]{2022MNRAS.515.3975A}. To guard against this possibility, we analysed the system further by splitting the data train. The primary goal was to examine the individual data halves for any candidate signals. If the baseline covered by any individual subset is appropriate for sampling the signals of interest, then any coherent signal should be detectable independently in $n$ number of data subsets, while the significance of detection depends on the size of the subset. \textcolor{black}{There is a caveat, though. It is probable that the stellar activity can manifest differently in the subsets and the ability of {\sc tweaks} to completely clean the time series could vary, leading to stronger signals in some cases, and weaker in others.}

\begin{table}

\caption{Bayesian evidence for models with and without {\sc scalpels} vectors used for stellar activity decorrelation. }
\vspace{0.1cm}
\begin{center}
\begin{tabular}{c c  c c }
\hline\hline                
Star ID& {\sc kima}   & {\sc kima}  + {\sc scalpels} & Bayes Factor \\ 
&(log$\cal{Z}$$_{1}$) &(log$\cal{Z}$$_{2}$)& \textcolor{black}{$\cal{Z}$$_{2}$/$\cal{Z}$$_{1}$}\\

\hline 
\vspace{0.1cm}
HD\,166620 & -457.29  & -453.65 &  \textcolor{black}{38.09} \\ 
\vspace{0.1cm}

HD\,144579 & -362.94& -357.56 &  \textcolor{black}{217.02} \\ 

\hline 
\label{Evidence}
\end{tabular}
\end{center}
\end{table}

Figure\,\ref{fig:raw-offset_2halves_HD144579} shows the FIPs obtained from the posteriors of individual half subsets. Contrary to what would be expected from any coherent planetary signal, 
neither
of the signals of interest were consistently found in 
both
of the two data subsets. Each of these signals could therefore be an artefact of the sampling pattern resulting from a particular localised subset of the observations or the existing data may not be adequate to confirm these detections.


To validate this finding, we also performed some injection and recovery tests by injecting a Keplerian with an orbital period of 210.28 d and an RV semi-amplitude of 60 cm\,s$^{-1}$ (comparable to the semi-amplitude of the 284 d signal). We then repeated the analysis and successfully recovered the injected Keplerian as independent marginal detections in individual half-data subsets (Figure\,\ref{fig:raw-offset_injected60cm}). An injected signal with a slightly larger semi-amplitude (80 cm\,s$^{-1}$) was consistently better detected in the individual subsets as shown in Figure\,\ref{fig:raw-offset_injected80cm}. We also experimented with injecting a short-period signal with P = 5.12 d which is closer to the 7.39 d signal and with an RV semi-amplitude $K$ = 60 cm\,s$^{-1}$. Unlike the long-period injected signal, this signal was only recovered in one of the half sets (Figure\,\ref{fig:raw-offset_injected5.1d60cm}). On the other hand an 80 cm\,s$^{-1}$ injected Keplerian was successfully recovered in individual subsets (Figure\,\ref{fig:raw-offset_injected5.1d80cm}). This suggests that $K$ = 60 cm\,s$^{-1}$ is below the detection threshold for the data partitioning test at P = 5.12 d. By the same token, it also suggests that the 7.39 d signal might not be entirely spurious. However, without confirmation in independent subsets of the data, we cannot be certain. 

This provides a pragmatic estimate of the detection limit in the HARPS-N data to be somewhere between 60 and 80 cm\,s$^{-1}$, which is consistent with the formal estimation of detection threshold presented in Section  \ref{subsec:detectionlimits}. In addition, these simulations show that the data in hand are not sufficient enough to conclude the origin of the 7.39-day and 280-day signals.


\subsection{Impact of stellar activity mitigation using {\sc scalpels}, in the model selection}

The difference in evidence (a.k.a Bayes factor) between a model with decorrelation performed against stellar activity using {\sc scalpels} U-vectors and a model uncorrected for stellar activity is given in Table \ref{Evidence}. The shape component comprising the corresponding U-vectors for individual stars are shown in orange colour in both panels of Figure\,\ref{fig:scalpels}. As anticipated for a star in 
a Maunder-minimum state
with no identifiable magnetic activity cycles, the shape component obtained for HD\,166620 looks essentially flat. However, {\sc scalpels} effectively deals with the focus intervention that occurred in the HARPS-N instrument, which is seen as an evident jump in the shape time series, at around BJD= 2456737. 
This intervention was planned as the FWHM in all stars observed by HARPS-N was showing a significant drift, likely due to a change of instrument focus with time. A component was removed that was supposed to stabilize the focus over time but was clearly not performing as expected. This fixed the FWHM drift issue, however, a significant jump in FWHM, likely due to a PSF change, was observed after this intervention.
This jump has been efficiently tracked by some of the {\sc scalpels}  basis vectors (Figure\,\ref{fig:HD166620corner} \& \ref{fig:HD144579corner}).

As is evident from the shape component in the lower panel of Figure\,\ref{fig:scalpels}, {\sc scalpels} 
identified a greater amplitude of shape-driven RV variation in HD\,144579 than in 
HD\,166620, 
owing to its greater intrinsic activity level. If we now take a look at Table \ref{Evidence}, we can see how the Bayes Factor  
reflects the impact of shape corrections on each star. The Bayes factor provides a measure of how strongly one model is supported over the other by the data.  We use the same Jeffreys' scale as a benchmark for our model evaluation.  In accordance with 
\citet{2008ConPh..49...71T}, we defined the threshold for "moderate" evidence at Bayes Factor = 12 and for "strong" evidence at Bayes Factor = 150. While {\sc scalpels} decorrelation improves the model by a factor of 38 for HD\,166620, HD\,144579 shows a significant improvement of 217 times.

\begin{figure}
    \centering
	
    \includegraphics[width=\columnwidth]{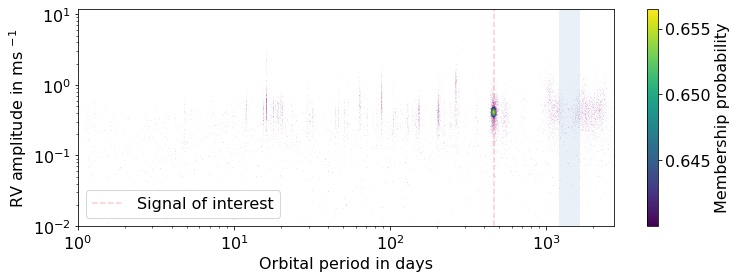}
    \includegraphics[width=\columnwidth]{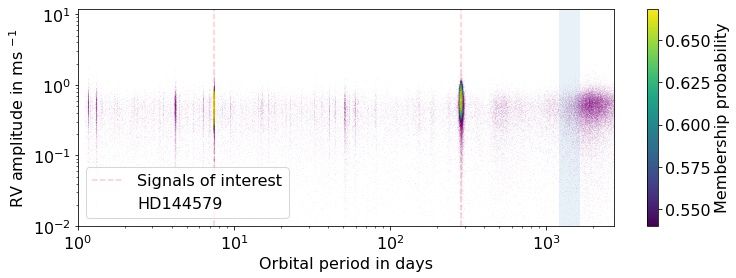}
	
    \caption{ The purple points represent the entire {\sc kima}  joint posterior semi-amplitudes plotted against the orbital periods. The green to yellow points represent the posterior samples that correspond to the foreground probability of the candidate signals as determined by a Gaussian mixture model. A point is more likely to be in the foreground if it is more yellow. Top: HD\,166620 Bottom: HD\,144579.}
    \label{fig:mixturemodel}
\end{figure}

\begin{figure*}
    \centering
	
    \includegraphics[width=1.5\columnwidth]{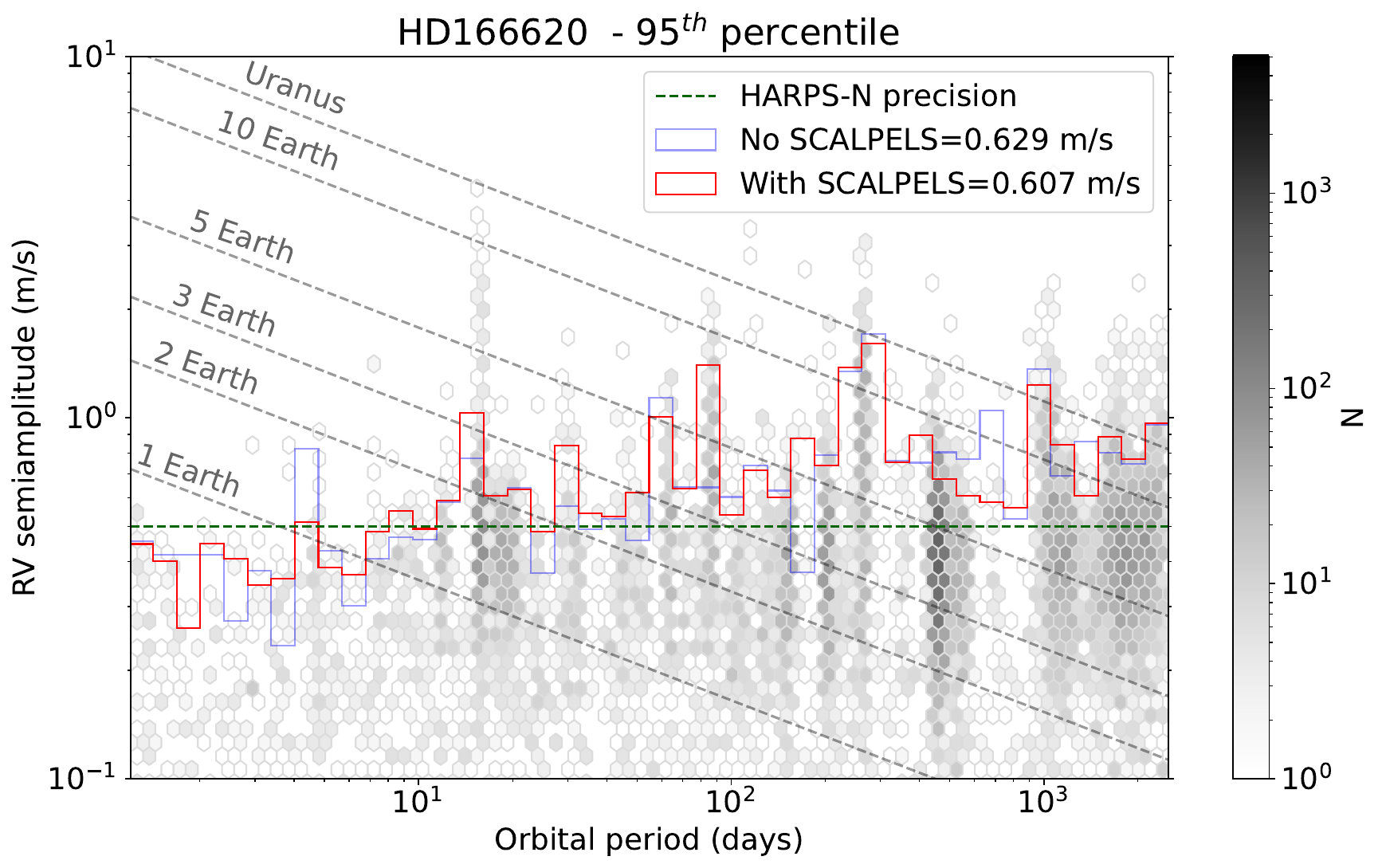}
    \includegraphics[width=1.5\columnwidth]{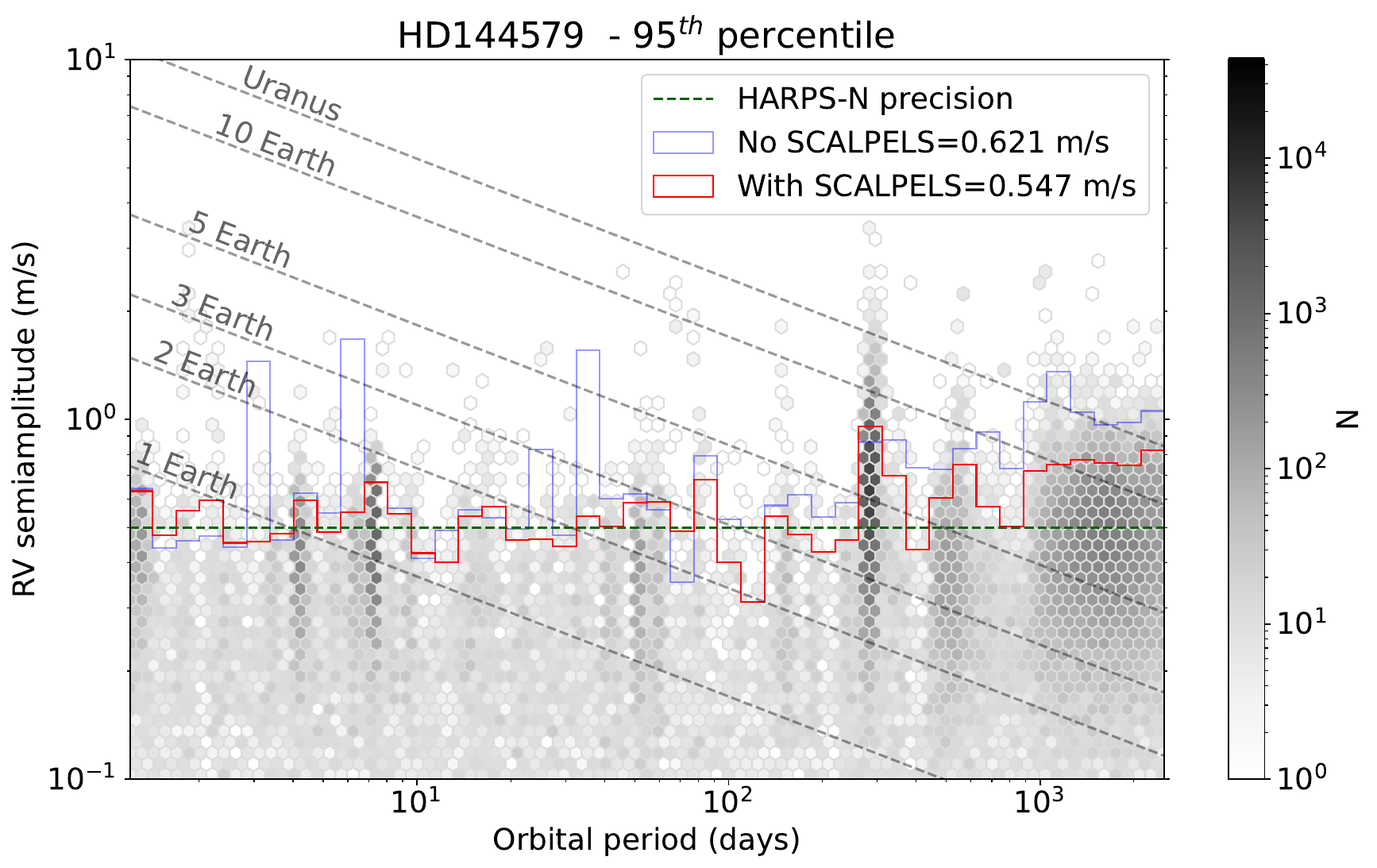}
	
    \caption{ A greyscale hexbin plot representing the density of posterior samples derived from the {\sc kima}+{\sc scalpels}  runs. Faded blue lines represent detection limits derived from a {\sc kima} run without any decorrelation against the scalpels basis vectors. The solid red line shows the detection limit computed from a {\sc kima} run corrected for stellar activity with {\sc scalpels}. The grey dashed lines are contours of constant planet mass ($M\sin i$) for comparison. }
    \label{fig:detectionlimits}
\end{figure*}

\begin{figure}
    \centering
	
    \includegraphics[width=1\columnwidth]{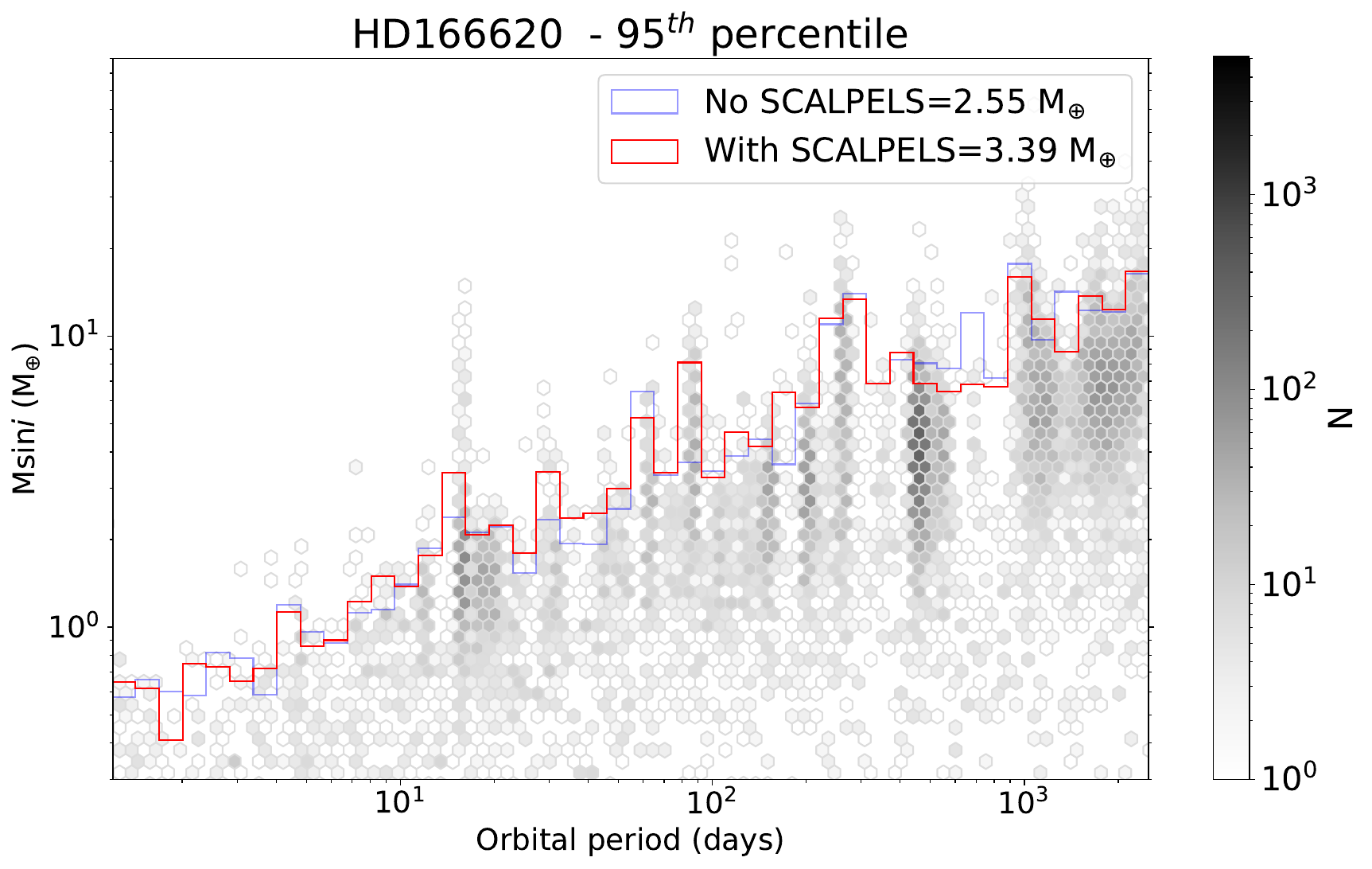}
    \includegraphics[width=1\columnwidth]{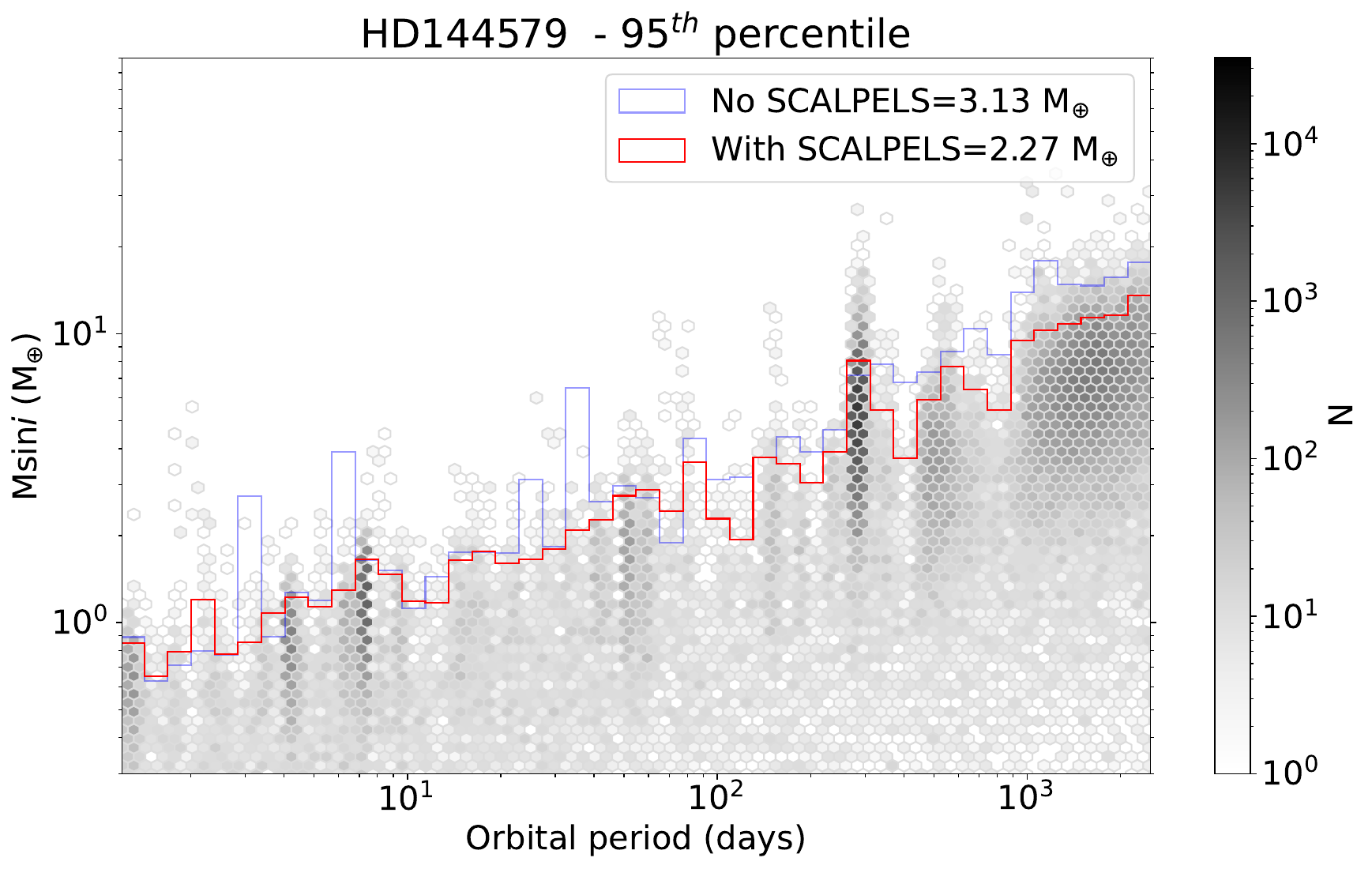}
	
    \caption{ Upper detection limits in the  $M\sin i$ versus orbital period space. The blue and red stepped histograms respectively show the 95$^{\rm th}$ percentile detection limit before and after decorrelating against {\sc scalpels} \textbf{U}-vectors. } 
    \label{fig:detectionlimitsMP}
\end{figure}

\section{Discussion}
\subsection{Detection limits}
\label{subsec:detectionlimits}
A distinct characteristic of a trans-dimensional nested sampler, such as {\sc kima} , is that it will create a map of all signals that are consistent with the data when forced to explore higher $N_{\rm p}$ than is explicitly detected. Since those suggested signals are still technically undetected, this posterior can be used to generate a detection threshold \citep{2022MNRAS.511.3571S}. To establish the detection limits in HD\,166620 and HD\,144579, we typically adhere to the process described in \citet{2022MNRAS.511.3571S}, where they use the Bayesian approach with KIMA to produce detection limits to answer the question ``what is compatible with the data?''. A well-sampled posterior is necessary for generating a reliable detection limit. The number of saves in {\sc kima}  was set to 200000 to obtain a minimum of 30000 effective posterior 
models.
A robust detection limit is usually obtained 
after
the removal of any candidate Keplerians present in the data. Here, we can directly compute the detection thresholds from the posteriors, since 
no valid detection can be established using the data presented.

The posterior is divided into log-spaced bins in the orbital period ($P$) to calculate the detection limit. Within each bin, the 95th percentile of the semi-amplitude ($K$) distribution is then evaluated. The findings of our analysis of the detection limits for the HD\,166620 and HD\,144579 are shown in the top and bottom panels of Figure\,\ref{fig:detectionlimits}. The grey scale hexbins display the posterior sample density from the {\sc kima}  + {\sc scalpels} trials on each target. The red lines enveloping the entire posterior space represent the calculated detection limit from a single run with decorrelation performed against the {\sc scalpels} U-vectors. The no-{\sc scalpels} version is used to compute the faded blue detection limit line.

Figures \ref{fig:raw-offset_FIP_HD166620} and \ref{fig:raw-offset_FIP_HD144579} depict the effect of {\sc scalpels} correction in HD\,166620 and HD\,144579. While the window function artefact at P$\sim$460 d in HD\,166620 appears well below the detection threshold, the corresponding signals in HD\,144579 at 7.39  and 284 d obstruct the improvement in the detection limit despite being decorrelated against {\sc scalpels} U-vectors.

\subsection{Probing RV signals in the sub-m/s regime}

Doing the stellar activity decorrelation in the wavelength domain using {\sc scalpels} also has an evident impact in bringing down the detection limit more towards the sub m s$^{-1}$ regime. In Figure\,\ref{fig:detectionlimits}, the detection limits calculated 
with and without {\sc scalpels} decorrelation
are shown. Although the impact varies slightly across various period regimes, {\sc scalpels} decorrelation allows for better probing in the low-mass regime in both stars. This is in line with a recent study by \citet{2022MNRAS.514.2259S}, in which they show how the activity cycle affects the detection efficiency of planets in the solar case. They found that the likelihood of finding a low-mass planet falls by an order of magnitude during the solar activity maximum. Therefore, they suggest a more efficient observational strategy based on a forecast of stellar activity that would enhance the identification of Earth counterparts and their analysis. Our study shows that we could detect smaller planets, when we correct for the stellar activity and instrumental shifts using {\sc scalpels}, despite the activity state of the target star. However, the level of magnetic activity in the individual stars is reflected in the improvement in the ability to detect low-mass planets on the application of adequate stellar activity correction.

While the detection limit does not show significant enhancement in HD\,166620, the mean detectability in HD\,144579 improves 
from 0.62 to 0.54 m s$^{-1}$
after correcting for the spectral-line shape changes. The most striking improvement is observed in the long-period regime, where the detection threshold consistently improves 
at periods longer than
300 days, reducing the detection threshold from $\sim 1.2$ m s$^{-1}$ to $\sim 0.6$ m s$^{-1}$ at periods around
1000 days. In the era of upcoming missions like PLATO and HARPS3, striving for Earth-like planets in an Earth-like orbit around a Sun-like star, these findings are of high importance.
It is evident from Figure\,\ref{fig:detectionlimits} that we are sensitive to a wide variety of Super-Earths (5-10\mearth) in the entire orbital period range ($P<2800$ d) spanned by the posterior space. Moreover, we can detect planets with masses as low as 1-3\mearth ($P<300$ d), and more massive planets with 5\mearth up to $P=1000$ d. The mean detection threshold computed after correcting for stellar activity falls around 60 cm s$^{-1}$ in HD\,166620 and 54\,cm s$^{-1}$ in HD\,144579, approaching the HARPS-N precision limit imposed by wavelength solution \citep{2021A&A...648A.103D}, which, without considering instrument intervention, should provide a similar long-term calibration precision for HARPS-N. 

Figure\,\ref{fig:detectionlimitsMP} displays the same detection limit figure translated into a $M\sin i$-orbital period space. It is now easily readable that under $P=10$ d, we are able to detect significantly lower-mass planets in HD\,166620 and HD\,144579 with  masses smaller than Earth. 

\subsection{Areas for further development}

Given that {\sc tweaks} does not completely model the rotational modulation of the star, there is room for further advancement. 
\citet{2022MNRAS.515.3975A} found that some shift-like patterns evade {\sc scalpels} analysis. They attribute the origin of these patterns to stellar rotation. We tried to use a GP to model any residual rotationally modulated signals. We were unable to proceed, as the typical Gaussian Process hyperparameters change considerably over the course of a cycle, making it more difficult to include in studies with long-term follow-ups like RPS. In addition, when we attempt to model the data using a single quasi-periodic GP, 
injected planetary signals with orbital periods longer than the active-region lifetime parameter are modelled as long-term trends, becoming attenuated.
A multi-variate GP 
using the {\sc scalpels} basis vectors as activity indicators could be used to optimally resolve this \citep[e.g.,][]{2015MNRAS.452.2269R,2022MNRAS.509..866B}. However, this is beyond the scope of the current investigation.


\section{Conclusions}

In this study, we report a thorough investigation of the HARPS-N spectroscopic data of bright K and G dwarfs HD\,166620 and HD\,144579, 
neither of which is known to host a planetary companion. We examined these stellar systems using {\sc tweaks}, 
that combines wavelength-domain and time-domain stellar activity mitigation using the {\sc kima} and {\sc scalpels}. The major objective was to search for planetary reflex-motion and investigate the impact of stellar activity mitigation on the capability of detecting low-mass planets, in stars with different levels of magnetic activity. We found no significant detections in either of the stars. We also ruled out the possibility of being misled by erroneous signals from sampling patterns by performing data splitting exercises and validating using injection recovery tests. The injection recovery tests also showed that the data in hand might be sufficient enough to conclude the origin of the 7.39-day and 280-day signals. The data partitioning test is something we would advise as a standard procedure to confirm the coherency of a signal before making any detections public. 
More rigorous coherency tests have also been proposed by \citet{2016MNRAS.458.2604G} and \citet{2022A&A...658A.177H}. 

Additionally, we provide an estimate for the detection limits on the HARPS-N radial velocity data using posteriors from {\sc kima} diffusive nested sampler.
We demonstrate the varied impact of stellar activity on the detection efficiency of planets in HD\,166620 which is in the Maunder minimum state and HD\,144579 which is moderately active.  The likelihood of finding a low-mass planet increases noticeably across a wide period range when the inherent star variability is corrected for using {\sc scalpels} \textbf{U}-vectors. The 54 cm s$^{-1}$ detection threshold achieved based on the aforementioned decorrelations brings us closer to the known calibration precision value provided by the HARPS-N instrument (50 cm s$^{-1}$) \citep{2021A&A...648A.103D}. 



\section*{Acknowledgements}

AAJ acknowledges the support from World leading St. Andrews Doctoral Scholarship. ACC and TGW acknowledge support from STFC consolidated grant numbers ST/R000824/1 and ST/V000861/1, and UKSA grant number ST/R003203/1. The results obtained in this study are based on the observations made with the Italian {\it Telescopio Nazionale
Galileo} (TNG) operated by the {\it Fundaci\'on Galileo Galilei} (FGG) of the
{\it Istituto Nazionale di Astrofisica} (INAF) at the
{\it  Observatorio del Roque de los Muchachos} (La Palma, Canary Islands, Spain). AAJ thanks Steven Saar for providing access to the 50 years of Keck and Mount Wilson data of HD\,166620, and Lars Buchhave for contributing to the estimation of robust stellar parameters for both stars. AAJ also acknowledges the data from HARPS-N project which was funded by Prodex programme of Swiss Space Office (SSO), the Harvard University Origin of Life Initiative (HUOLI), Scottish Universities Physics Alliance (SUPA),  University of Geneva, the Smithsonian Astrophysical Observatory (SAO), and the Italian National Astrophysical Institute (INAF), University of St. Andrews, Queen\'s University Belfast, and the University of Edinburgh. J.P.F acknowledges Funda\c{c}\~ao para a Ci\^encia e a Tecnologia (FCT, Portugal) for the research grants UIDB/04434/2020 and UIDP/04434/2020 and by POCH/FSE (EC) through the grant EXPL/FIS-AST/0615/2021. J.P.F. is supported in the form of a work contract funded by national funds through FCT with reference DL57/2016/CP1364/CT0005. KR is grateful for support from UK STFC via consolidated grant ST/V000594/1. R.D.H. is funded by UK Science and Technology Facilities Council (STFC)'s Ernest Rutherford Fellowship (grant number ST/V004735/1. This project has received funding from European Research Council (ERC) under  European Union\'s Horizon 2020 research and innovation programme (grant agreement SCORE No 851555). This work has been carried out within the framework of National Centre of Competence in Research PlanetS supported by  Swiss National Science Foundation under grants 51NF40\_182901 and 51NF40\_205606. This work has made use of data from European Space Agency (ESA) mission {\it Gaia} (\url{https://www.cosmos.esa.int/gaia}), processed by {\it Gaia} Data Processing and Analysis Consortium (DPAC, 
\url{https://www.cosmos.esa.int/web/gaia/dpac/consortium}). Funding for the DPAC has been provided by national institutions, in particular, the institutions participating in the {\it Gaia} Multilateral Agreement. The authors acknowledge the financial support of the SNSF. In order to meet institutional and research funder open access requirements, any accepted manuscript arising shall be open access under a Creative Commons Attribution (CC BY) reuse licence with zero embargoes.
\section*{Data Availability}

The HARPS-N data products used for the analyses are  from the Data \& Analysis Center for Exoplanets (DACE) database  dedicated to the visualization, exchange and analysis of extrasolar planets' data. The {\sc python} codes and notebooks used to generate the results and diagrams in this paper will be made available through the University of St Andrews Research Portal.






\begin{appendix}
    
\textcolor{black}{\section{Instrumental zero-point correction}}

\textcolor{black}{Table \ref{tab:offsetsample} shows the information about the 12 stars included in the sample set used to estimate the RV zero-point between the cryostat warmups described in Section \ref{subsec:zeropoint}. The epochs of cryostat warmups and the consequently obtained RV zero points between the intervals of these instrumental interventions are given in Table \ref{tab:RVoffsets}.}

\begin{table}
\caption{RPS targets used in the sample dataset to monitor the 1400-day periodicity and calculate the zero points. Note: HD\,166620 is taken out from the list when estimating zero points for itself. Similarly, for HD\,144579 as well.} 
    \centering
    \begin{tabular}{c c c c}
    \hline\hline
    Star & No. of observations & After nightly binning & Time span (days) \\ [0.2ex]
    \hline 
    HD\,10476 & 1062 & 257 & 4108 \\
    HD\,122064 & 709 & 280 & 4187 \\ 
    HD\,127334& 1432 & 388 & 4023\\ 
    HD\,128165 & 473 & 268 & 3674\\
    HD\,144579 & 888 &  257 & 3646\\ 
    HD\,158633 & 626 & 167 & 3301\\ 
    HD\,166620& 947  & 318 & 4080\\ 
    HD 32147  & 505 & 139 & 4091\\
    HD 3651&   720  & 154 & 3558\\ 
    HD 4628   & 1540& 379 & 4099\\
    HD 62613 & 527 & 157  & 3976\\ 
    Sun & 857&&\\
    \end{tabular}
    \label{tab:offsetsample}
\end{table}

\begin{table}
\caption{\textcolor{black}{BJD boundaries and RV zero-point offsets calculated from the full sample including both HD\,166620 and HD\,144579. The `nan' values for offsets represent the intervals where no observations were obtained.}} 
    \centering
    \begin{tabular}{c c }
    \hline\hline
    Date of intervention  & RV zeropoints  \\ [0.2ex]
    \hline 
    2456738.50000 & -0.2805  \\
    2456826.50000 & -0.9587 \\ 
    2456947.50000 &  -0.7384 \\
    2457056.50000 & -1.6168\\
    2457072.50000 & nan\\
    2457076.50000 & nan\\
    2457161.50000 & -0.2981\\
    2457308.50000 & nan\\
    2457478.50000 & -0.1927\\
    2457687.50000 & 0.4363\\
    2457854.50000 & -0.0994\\
    2458071.50000 &  0.1848\\
    2458231.50000 & 0.4039\\
    2458412.50000 & -0.2475\\
    2458554.50000 & -0.4048\\
    2458683.50000 & -0.5337\\
    2458839.50000 & 0.2337\\
    2458997.50000 & 0.0001\\
    2459049.50000 & 0.6544\\
    2459180.50000 & 0.7807\\
    2459316.50000 & 1.0554\\
    2459390.50000 & -0.0803\\
    \end{tabular}
    \label{tab:RVoffsets}
\end{table}

\textcolor{black}{\section{Injection \& recovery tests}}

\begin{figure}
    \centering
	
    \includegraphics[width=1\columnwidth]{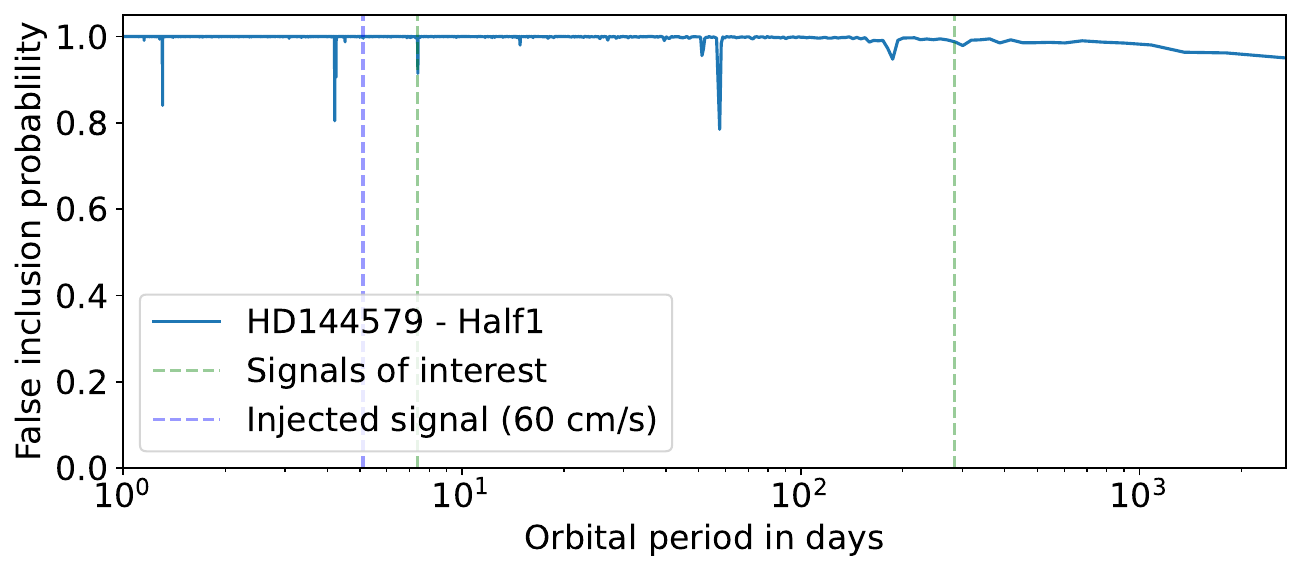}
    \includegraphics[width=1\columnwidth]{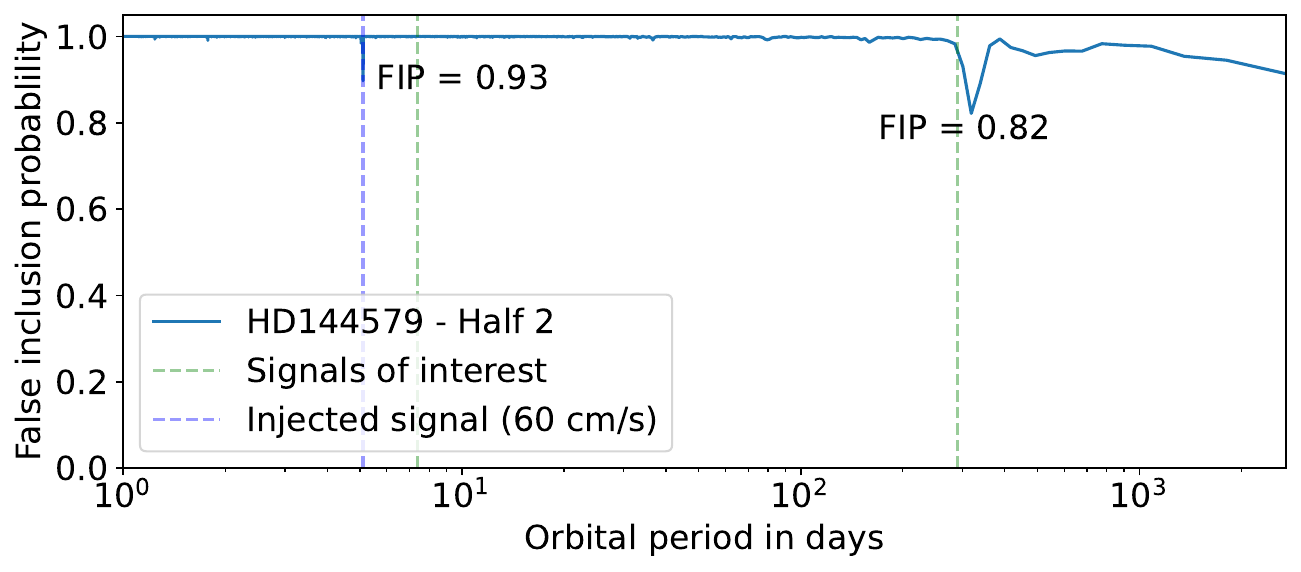}
	
    \caption{ A Keplerian with P=5.12 d and $K$=0.6 m s$^{-1}$ is injected into the data.  The FIP diagrams obtained for individual half subsets of the HD\,144579 data are shown in the two panels. The injected Keplerian is only detected in one of the data sets.}
    \label{fig:raw-offset_injected5.1d60cm}
\end{figure}

\begin{figure}
    \centering
	
    \includegraphics[width=1\columnwidth]{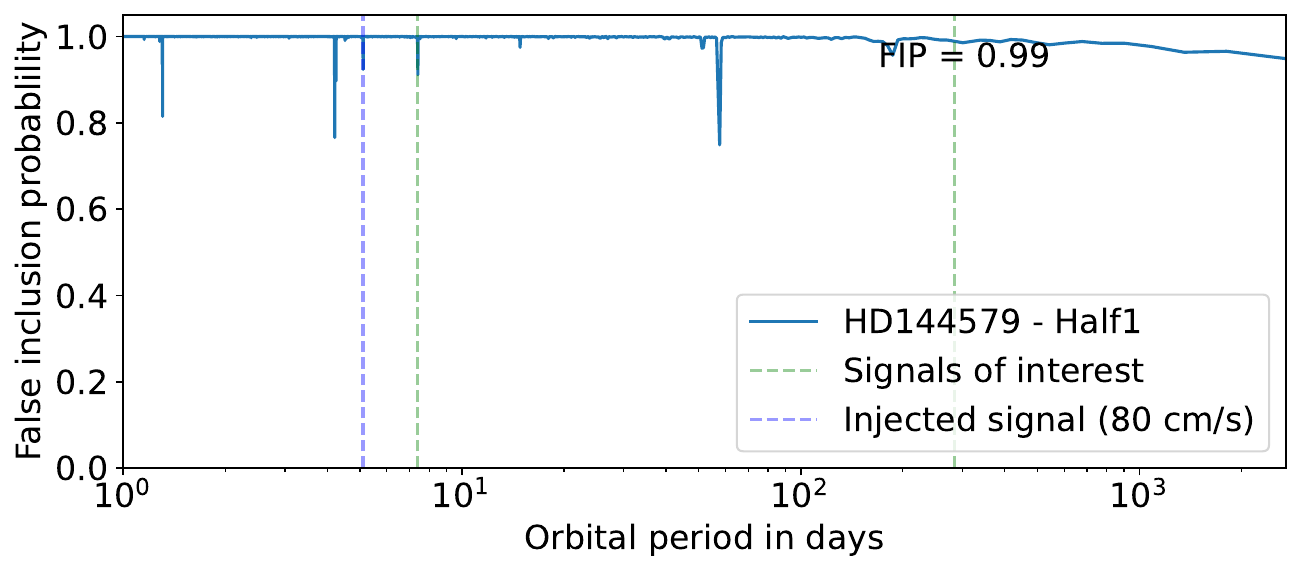}
    \includegraphics[width=1\columnwidth]{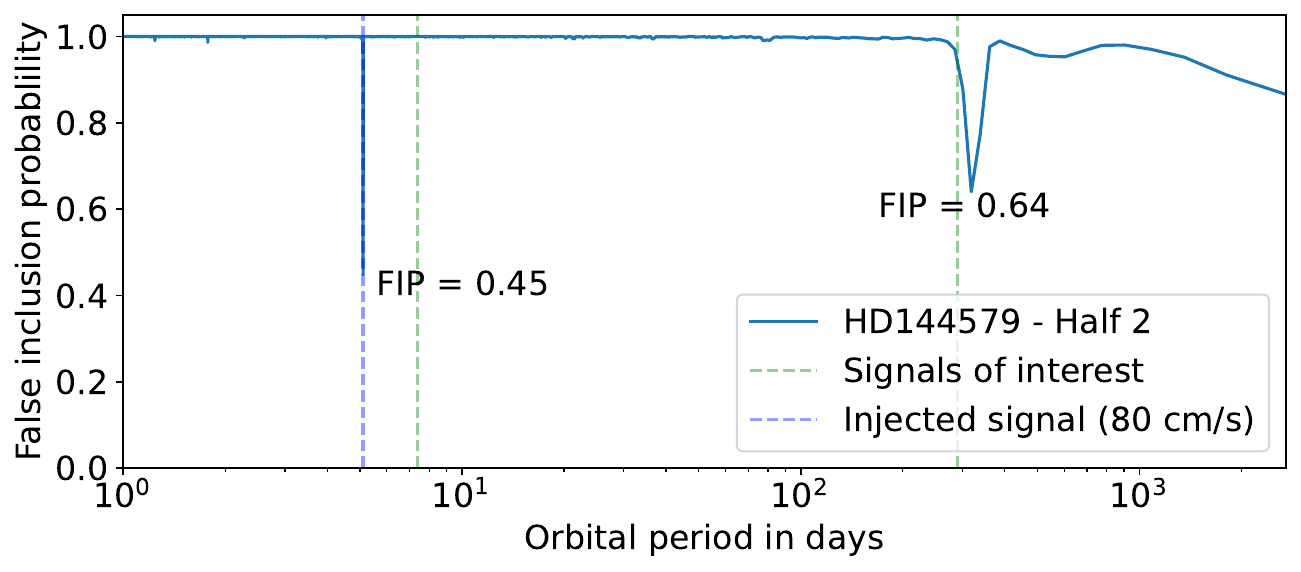}
	
    \caption{ A Keplerian with P=5.12 d and $K$=0.8 m s$^{-1}$ is injected into the data.  The FIP diagrams obtained for individual half subsets of the HD\,144579 data are shown in the two panels. The injected Keplerian is now detected in both data sets independently, strongly in one half and marginally in the other half. \textcolor{black}{This can be due to the cross talk of the injected signal between the nearby 7.34 day signal and its strong 1-day alias possibly arising from the sampling pattern of the first half data set.}}
    \label{fig:raw-offset_injected5.1d80cm}

\end{figure}

\begin{figure}
    \centering
	\includegraphics[width=1\columnwidth]{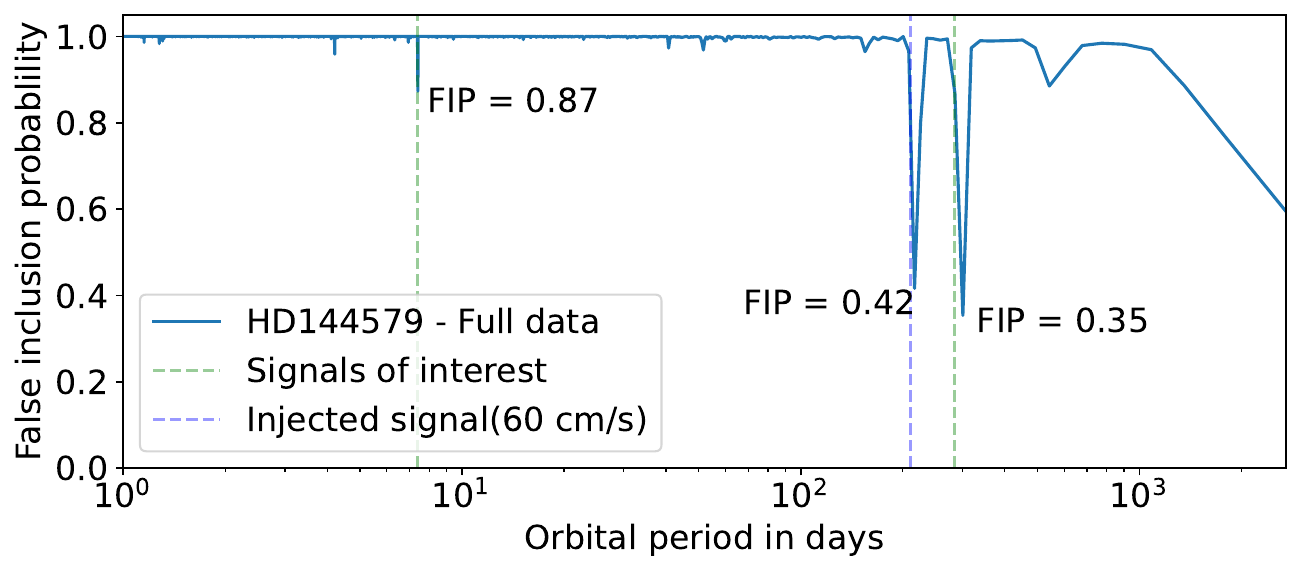}
    \includegraphics[width=1\columnwidth]{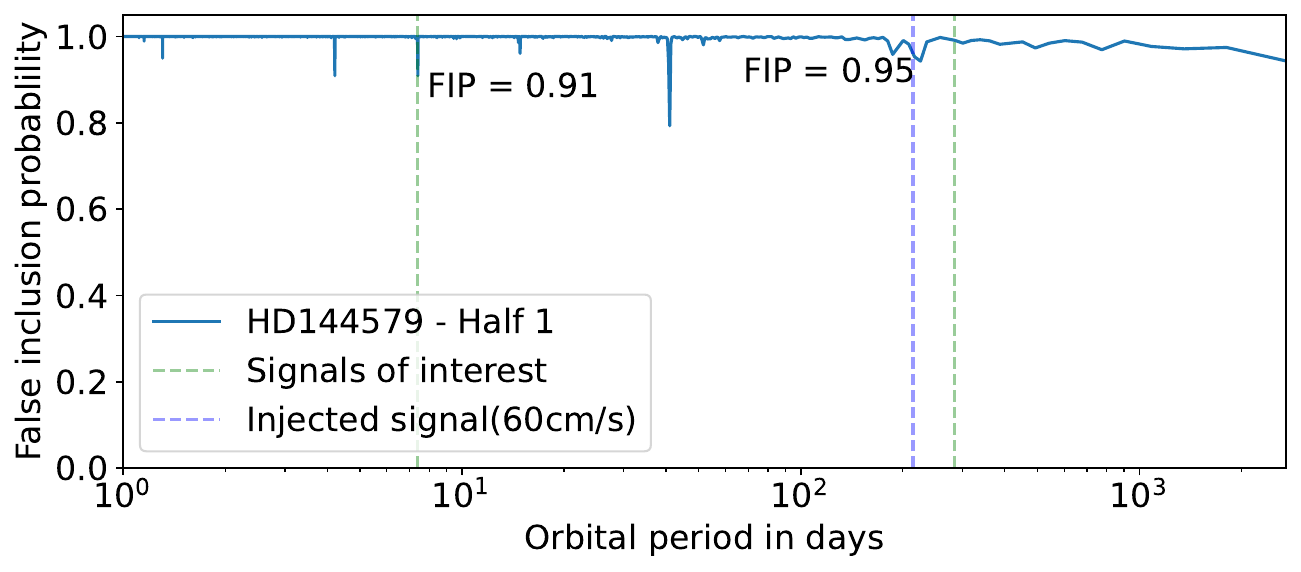}
    \includegraphics[width=1\columnwidth]{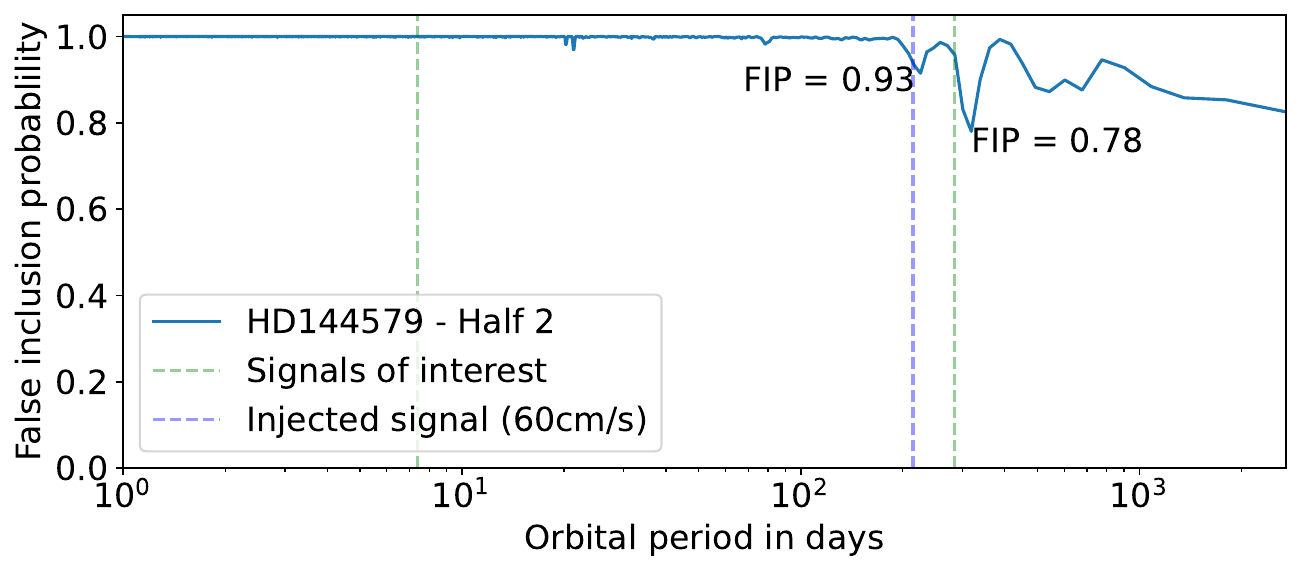}
	
    \caption{ A Keplerian with P=210.28 d and $K$=0.6 m s$^{-1}$ is injected to the data.  The observations are then divided into two subsets to investigate how a coherent signal appears in the individual halves. The FIP diagrams obtained for individual half subsets of the HD\,144579 data are shown in the two panels. Unlike the signals of interest at P= 7.39 and 284 d, the injected Kepelerian is marginally detected in both data sets independently.}
    \label{fig:raw-offset_injected60cm}
\end{figure}

\begin{figure}
    \centering
	\includegraphics[width=1\columnwidth]{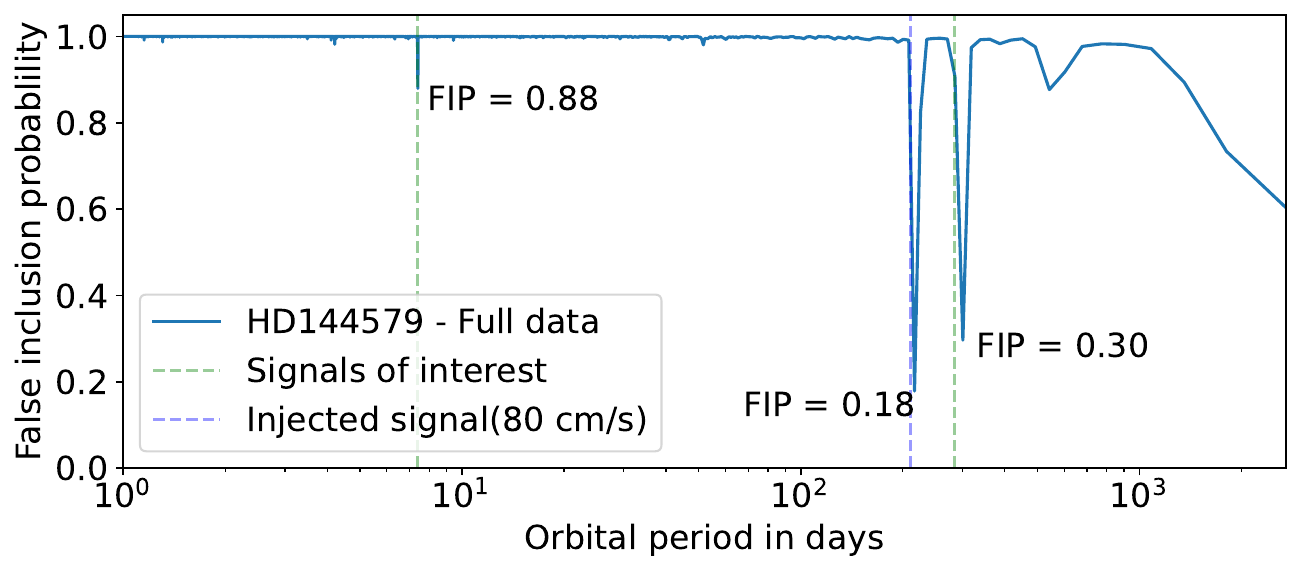}
    \includegraphics[width=1\columnwidth]{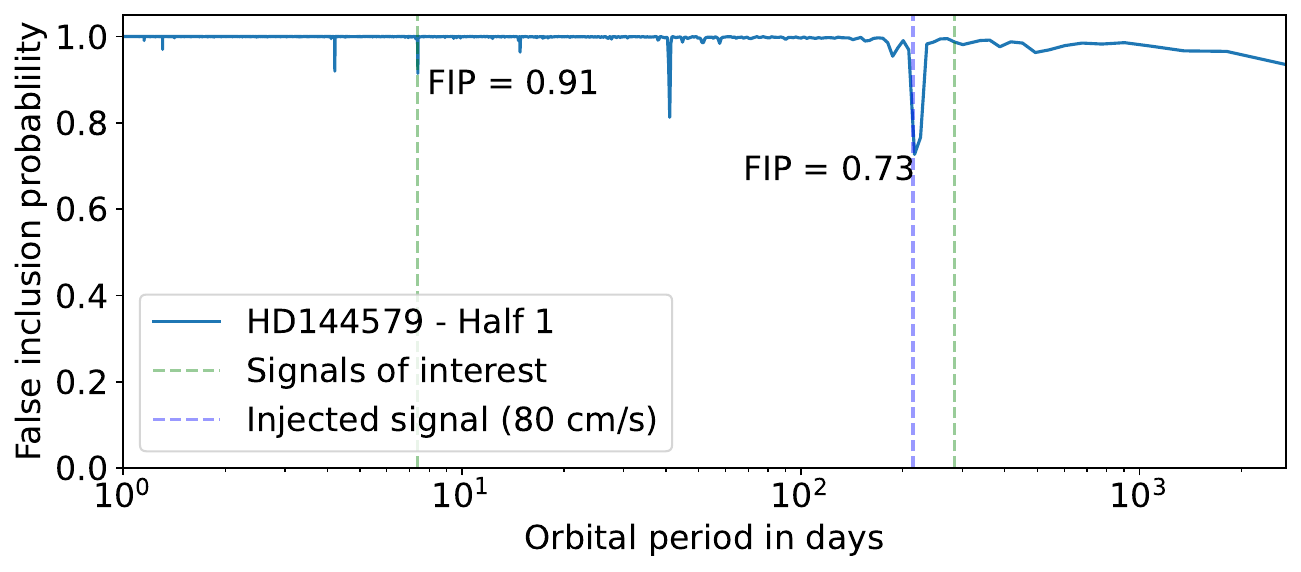}
    \includegraphics[width=1\columnwidth]{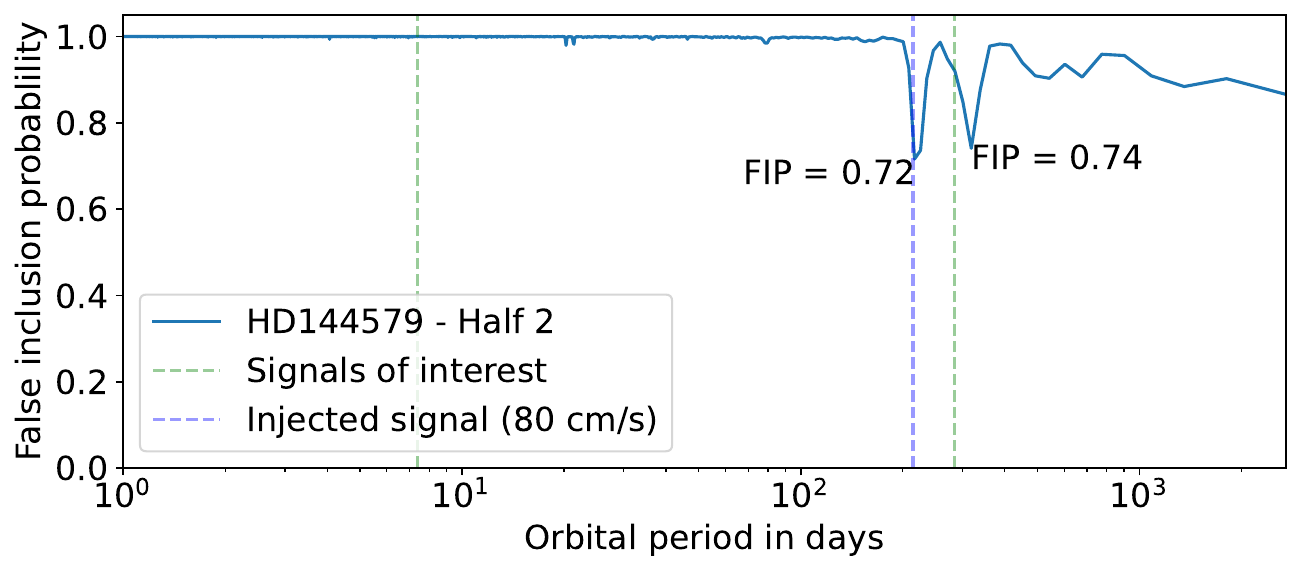}
	
    \caption{ A Keplerian with P=210.28 d and $K$=0.8 m s$^{-1}$ is injected to the data.  The FIP diagrams obtained for individual half subsets of the HD\,144579 data are shown in the two panels. The injected Keplerian is now \textcolor{black}{better} detected in both data sets independently. }
    \label{fig:raw-offset_injected80cm}
\end{figure}

\textcolor{black}{The FIP periodograms from the injection recovery tests, performed with injected Keplerians of varied orbital periods and RV semi-amplitudes are shown in Figures \ref{fig:raw-offset_injected5.1d60cm}, \ref{fig:raw-offset_injected5.1d80cm}, \ref{fig:raw-offset_injected60cm} and \ref{fig:raw-offset_injected80cm}. }
As discussed in Section \ref{subsec:HD144579}, we performed 4 injection and recovery tests.\\
\textcolor{black}{1) Injecting a signal with $K$=60 cm/s and P = 5.12 days (Figure\,\ref{fig:raw-offset_injected5.1d60cm}), closer to the signal of interest at P=7.34 days.\\}
\textcolor{black}{2) Injecting a signal with $K$=80 cm/s and P = 5.12 days (Figure\,\ref{fig:raw-offset_injected5.1d80cm})\\}
\textcolor{black}{3) Injecting a signal with $K$=60 cm/s and P = 210.28 days (Figure\,\ref{fig:raw-offset_injected60cm}), closer to the second signal of interest at P=285 days.\\}
\textcolor{black}{4) Injecting a signal with $K$=80 cm/s and P = 210.28 days (Figure\,\ref{fig:raw-offset_injected80cm}).}\\
\textcolor{black}{These injection \& recovery tests helped us to validate the conclusions about the incoherency of the signals of interest, as the injected signals appeared independently in individual data halves (weak though in the case of 60 cm/s signals), unlike the SOIs. The strength of detection as shown by the FIPs (in \ref{fig:raw-offset_injected60cm} and \ref{fig:raw-offset_injected80cm}) improved dramatically from 0.42 to 0.18 when the semi-amplitude of the injected signal was increased from 60 cm/s to 80 cm/s, which aided us in estimating the pragmatic detection limits to lie between 60 and 80 cm/s. }

\textcolor{black}{The FIP appears undoubtedly higher in the individual halves than the full data set (in both 60 cm/s and 80 cm/s cases). However, the idea here was to investigate if the injected signals show up independently in individual data halves with robustly defined orbital periods. The notable differences in the FIP in the full data also point towards the fact that doubling the number of observations improves the significance of detection drastically. This will also be important in planning observations in order to make good use of telescope time.}

\begin{figure}
    \centering
	
    \includegraphics[width=1\columnwidth]{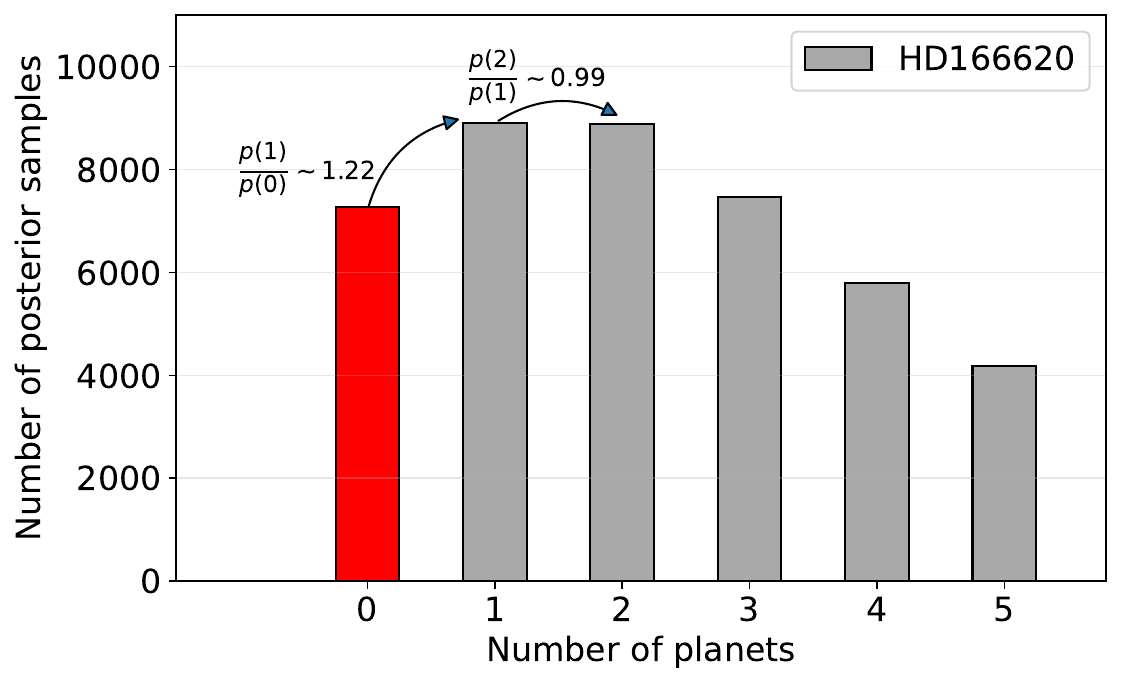}

    \caption{ The posterior distribution for HD\,166620, for the number of planets Np. The counts are the number of posterior samples in trial models with a specific number of planets. The probability ratios between models with 0,1 and 2 planets are shown. The posterior distribution suggests that there are likely to be `zero' planets present. }
    \label{fig:raw-offset_HD166620hist}
\end{figure}

\begin{figure}
    \centering
	
    \includegraphics[width=1\columnwidth]{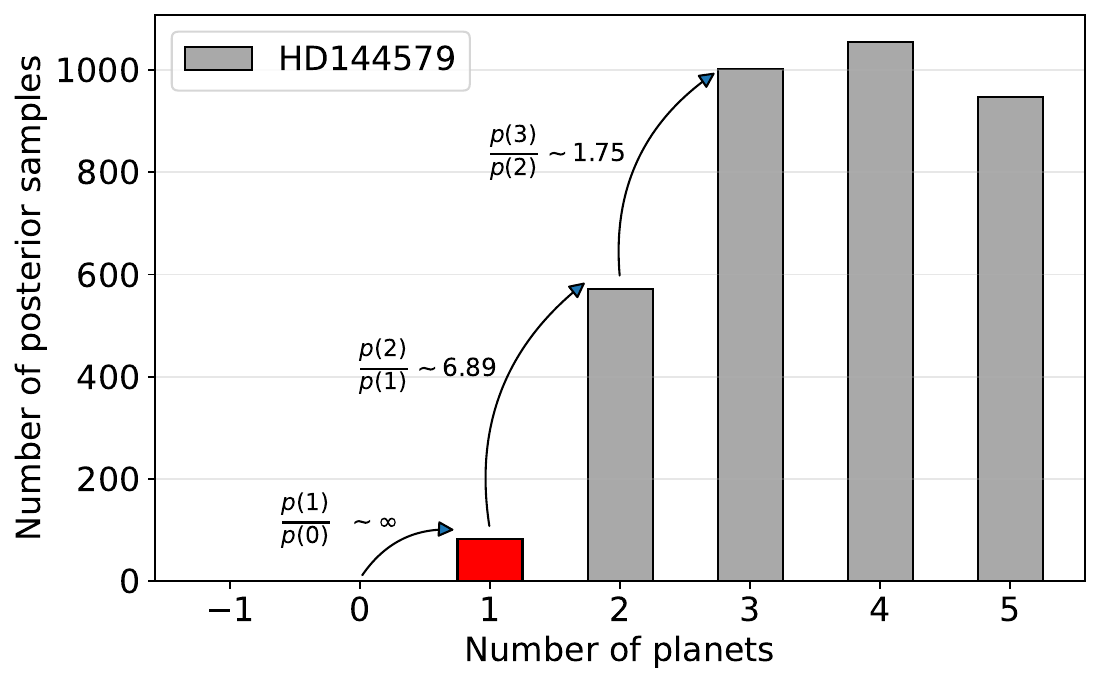}

    \caption{ A Keplerian with P=210.28 d and $K$=0.8 m s$^{-1}$ is injected to the data. The ratio of probabilities for successive values of N$_{\rm p}$ now favours N$_{\rm p}$=1.}
    \label{fig:raw-offset_injected80cmhist}
\end{figure}

\begin{figure}
	
    \includegraphics[width=1\columnwidth]{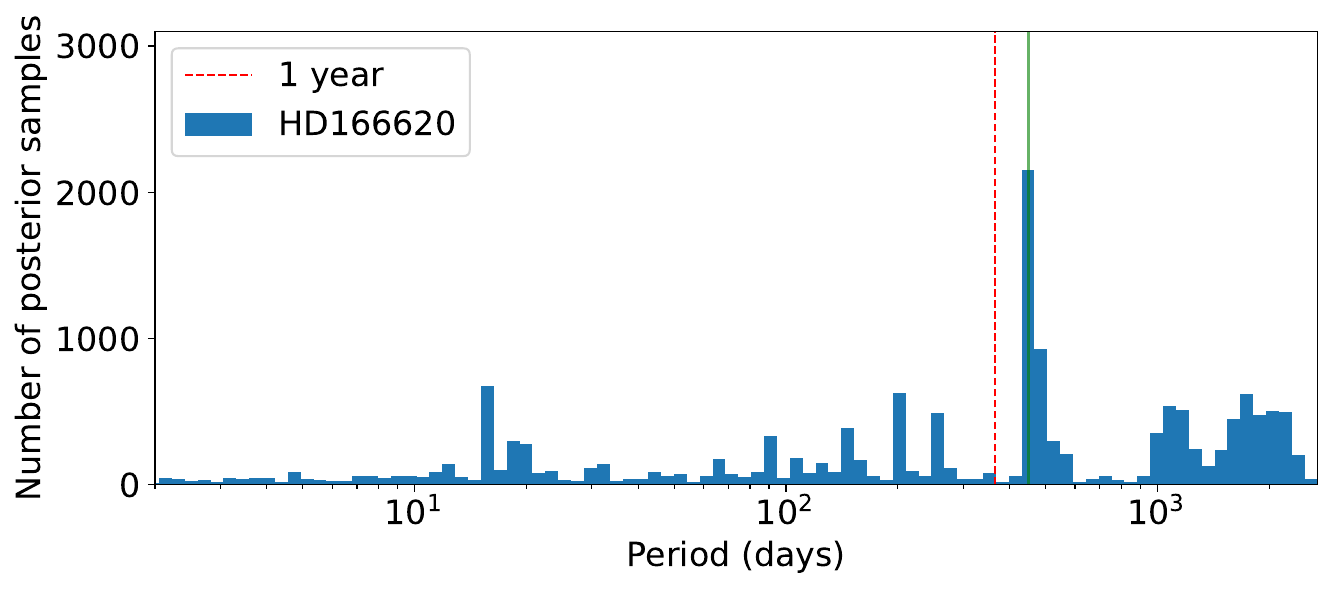}
    \includegraphics[width=1\columnwidth]{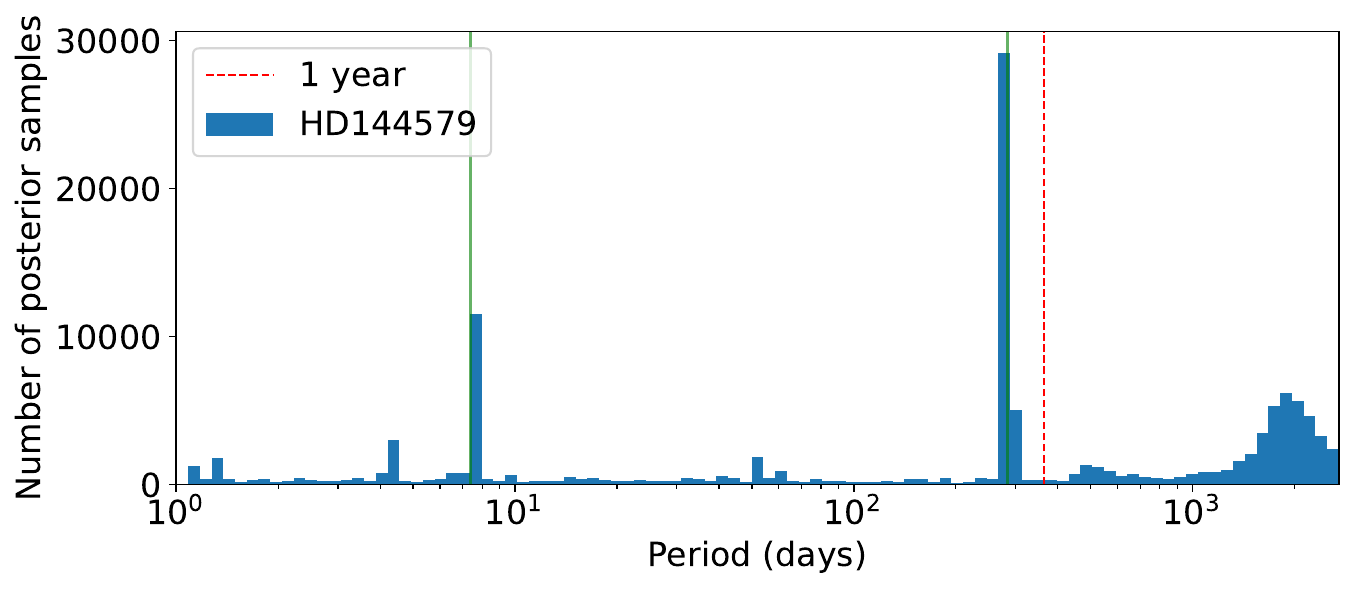}
	
    \caption{ Histograms showing the number of independent posterior samples for each star. The signals of interest are marked with green vertical lines.}
    \label{fig:raw-offset_hist}
\end{figure}

\textcolor{black}{\section{{\sc kima} planet models}}
\textcolor{black}{The {\sc kima} transdimensional nested sampling algorithm also provides a posterior distribution of the number of planets favoured by the data based on the ratios of Bayes Factor (or evidence) between successive models with different numbers of keplerians. A model with $N_{\rm p}$ planets is considered to be the most favoured (strongly detected) when this ratio $\cal{Z}$$_{N_{\rm p}}$/$\cal{Z}$$_{N_{\rm p-1}} > 150$, and moderately detected when > 38, based on the Jeffreys criteria. Figure\,\ref{fig:raw-offset_HD166620hist} shows this representation for HD\,166620, while the complementary Figure for HD\,144579 is discussed in Figure\,\ref{fig:nphist}, both of which show the supported zero-planet models. Figure\,\ref{fig:raw-offset_injected80cmhist} on the other hand shows the favoured `1-planet' model on the injection of the 80 cm/s signal in HD\,144579, pointing towards the detection threshold obtained in  Section \ref{subsec:detectionlimits}.}

\textcolor{black}{\section{{\sc scalpels} \textbf{U}-vectors \& activity indicators}}

\begin{figure*}
    \centering
    \includegraphics[width=2\columnwidth]{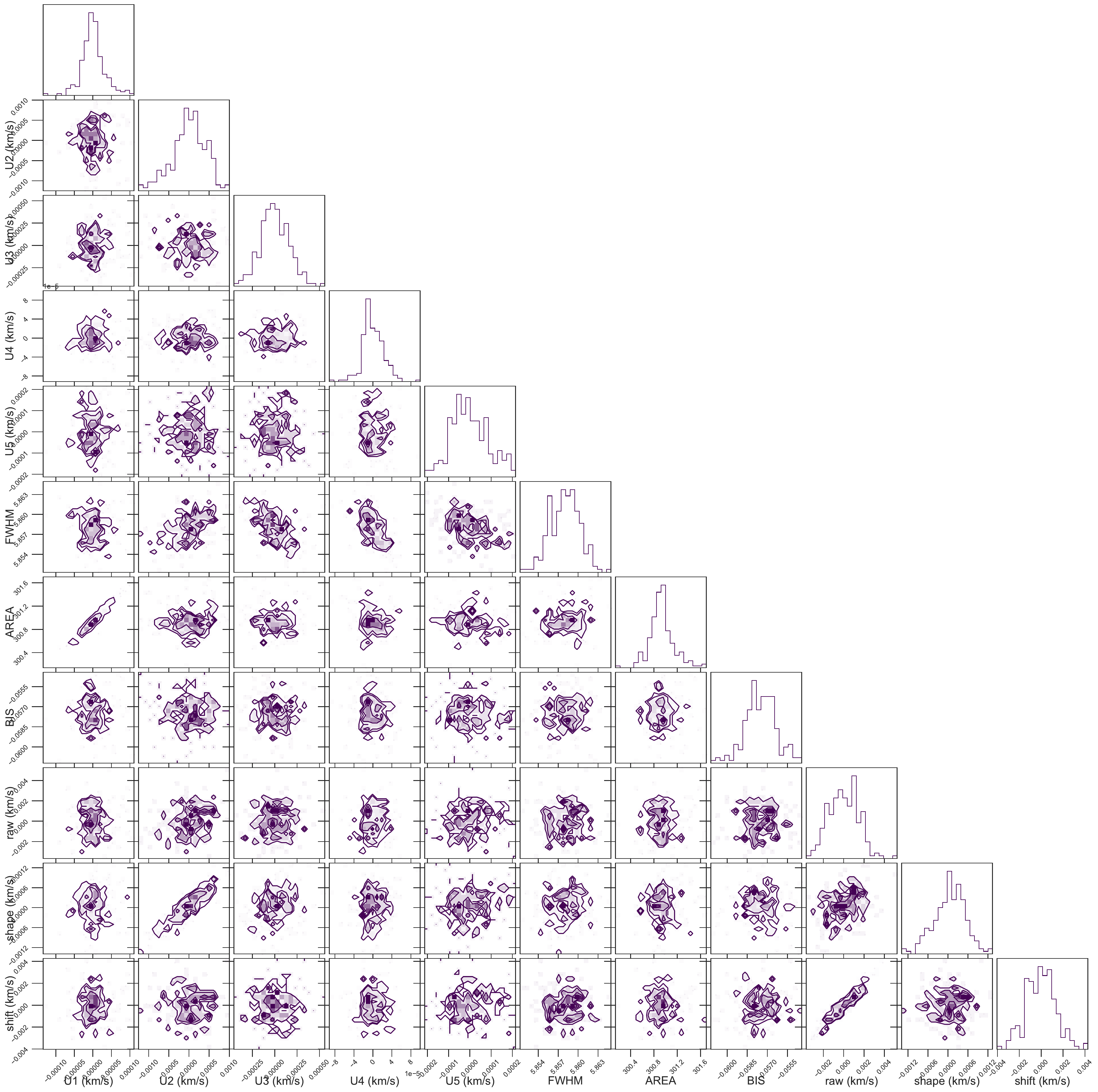}
    \caption{\textcolor{black}{The correlations plots for HD144579 showing the leading \textbf{U}-vector components of the residual CCF of RV time series and the traditional activity indicators such as FWHM, area and Bisector span. \textbf{U$_{2}$} is the prominent contributor to the entire shape component which shows a weak correlation with the FWHM, while the \textbf{U$_{1}$} component is strongly correlated with the area. \textbf{U$_{3}$} shows an anticorrelation with the FWHM.}}
    
    \label{fig:HD144579corner}
\end{figure*}

\begin{figure*}
    \centering
    \includegraphics[width=2\columnwidth]{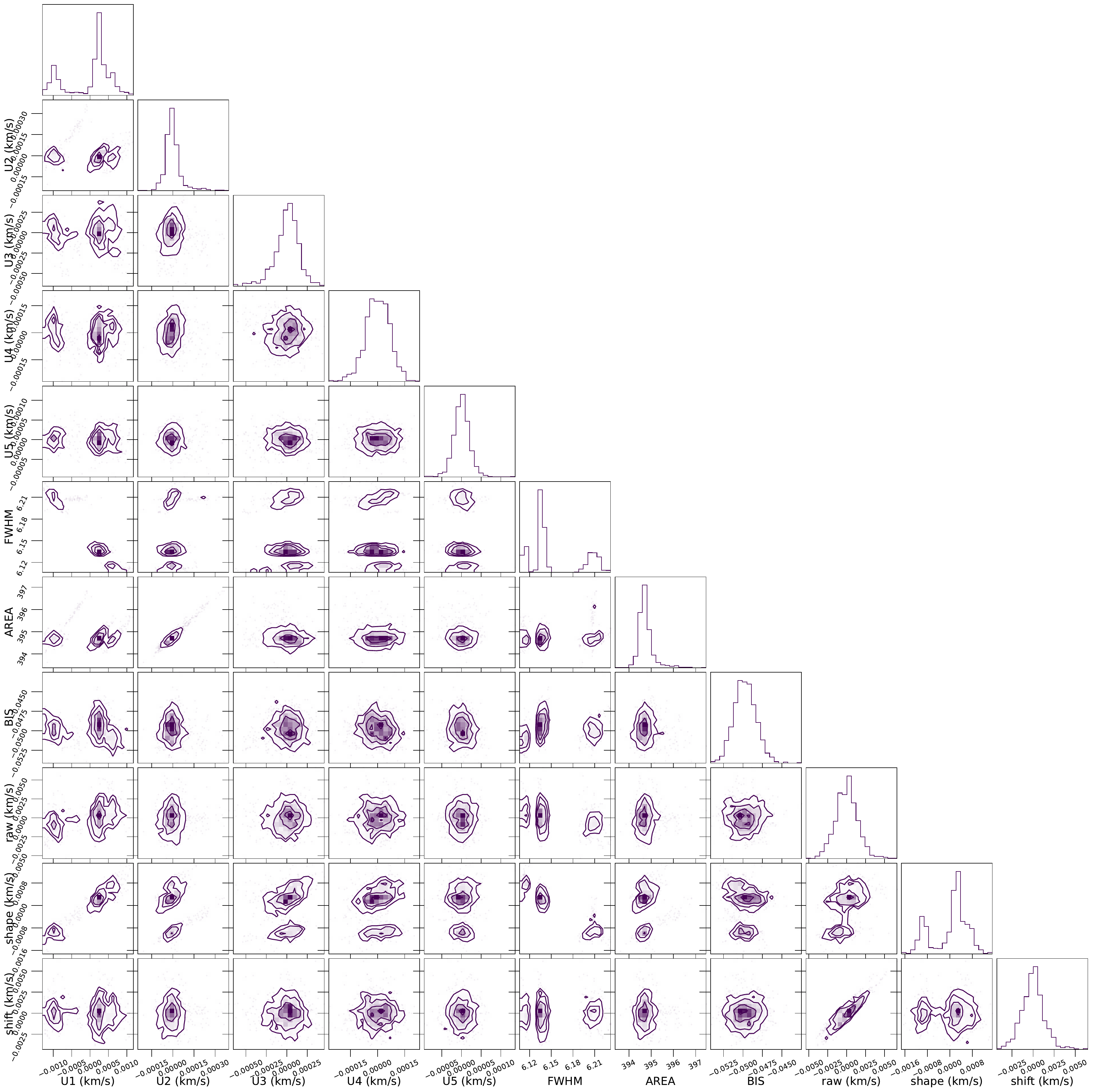}
    \caption{\textcolor{black}{The correlations plots for HD166620 showing the leading \textbf{U}-vector components of the residual CCF of RV time series and the traditional activity indicators such as FWHM, area and Bisector span. \textbf{U$_{1}$} is the prominent contributor to the entire shape component which shows a strong anti-correlation with the FWHM, while the \textbf{U$_{2}$} component is strongly correlated with the FWHM.}}
    
    \label{fig:HD166620corner}
\end{figure*}

\textcolor{black}{We also examined the first 5 leading principal components that contributed significantly in modelling the stellar activity (orange timeseries shown in Figure\,\ref{fig:scalpels}),
to understand how efficiently the Principal Component Analysis of
the ACF correlate with the known activity proxies such as FWHM, area and Bisector span. The corner plots showing the correlation between the {\sc scalpels} \textbf{U}-vectors and traditional activity indicators are shown for HD\,166620 and HD\,144579 in Figure\,\ref{fig:HD166620corner} and \ref{fig:HD144579corner} respectively.}

\textcolor{black}{In HD\,166620, the first and second principal components (\textbf{U}$_1$ \& \textbf{U}$_2$) effectively trace down the FWHM and area of the CCF respectively.
In HD\,144579, the first principal component (\textbf{U}$_1$) exhibits a strong resemblance to the variability of CCF area. Another correlation is
found between the second principal component((\textbf{U}$_2$) ) and the FWHM.
This indicates that the changes in the width of the profile are also reliably
considered while modelling the shape-driven component.}

\end{appendix}


\label{lastpage}
\end{document}